\documentclass[dvips]{acta}
\usepackage{supertabular,lscape,epsfig}
\usepackage{amssymb}
\usepackage{amsmath}
\usepackage{multirow}
\usepackage{float}
\mathchardef\mhyphen="2D
\usepackage{placeins}
\DeclareSymbolFont{ppa}{OT1}{ppl}{m}{it}
\DeclareMathSymbol{\vv}{\mathalpha}{ppa}{'166}

\SetPages{0}{0}

\SetVol{75}{2025}

\begin{document}
\newcommand\pvalue{\mathop{p\mhyphen {\rm value}}}
\newcommand{\TabApp}[2]{\begin{center}\parbox[t]{#1}{\centerline{
  {\bf Appendix}}
  \vskip2mm
  \centerline{\small {\spaceskip 2pt plus 1pt minus 1pt T a b l e}
  \refstepcounter{table}\thetable}
  \vskip2mm
  \centerline{\footnotesize #2}}
  \vskip3mm
\end{center}}

\newcommand{\TabCapp}[2]{\begin{center}\parbox[t]{#1}{\centerline{
  \small {\spaceskip 2pt plus 1pt minus 1pt T a b l e}
  \refstepcounter{table}\thetable}
  \vskip2mm
  \centerline{\footnotesize #2}}
  \vskip3mm
\end{center}}

\newcommand{\TTabCap}[3]{\begin{center}\parbox[t]{#1}{\centerline{
  \small {\spaceskip 2pt plus 1pt minus 1pt T a b l e}
  \refstepcounter{table}\thetable}
  \vskip2mm
  \centerline{\footnotesize #2}
  \centerline{\footnotesize #3}}
  \vskip1mm
\end{center}}

\newcommand{\MakeTableH}[4]{\begin{table}[H]\TabCap{#2}{#3}
  \begin{center} \TableFont \begin{tabular}{#1} #4 
  \end{tabular}\end{center}\end{table}}

\newcommand{\MakeTableApp}[4]{\begin{table}[p]\TabApp{#2}{#3}
  \begin{center} \TableFont \begin{tabular}{#1} #4 
  \end{tabular}\end{center}\end{table}}

\newcommand{\MakeTableSepp}[4]{\begin{table}[p]\TabCapp{#2}{#3}
  \begin{center} \TableFont \begin{tabular}{#1} #4 
  \end{tabular}\end{center}\end{table}}

\newcommand{\MakeTableee}[4]{\begin{table}[htb]\TabCapp{#2}{#3}
  \begin{center} \TableFont \begin{tabular}{#1} #4
  \end{tabular}\end{center}\end{table}}

\newcommand{\MakeTablee}[5]{\begin{table}[htb]\TTabCap{#2}{#3}{#4}
  \begin{center} \TableFont \begin{tabular}{#1} #5 
  \end{tabular}\end{center}\end{table}}

\newcommand{\MakeTableHH}[4]{\begin{table}[H]\TabCapp{#2}{#3}
  \begin{center} \TableFont \begin{tabular}{#1} #4 
  \end{tabular}\end{center}\end{table}}

\newfont{\bb}{ptmbi8t at 12pt}
\newfont{\bbb}{cmbxti10}
\newfont{\bbbb}{cmbxti10 at 9pt}
\newcommand{\uprule}{\rule{0pt}{2.5ex}}
\newcommand{\douprule}{\rule[-2ex]{0pt}{4.5ex}}
\newcommand{\dorule}{\rule[-2ex]{0pt}{2ex}}
\def\thefootnote{\fnsymbol{footnote}}
\begin{Titlepage}
\Title{Automated Detection and Modeling of Binary Microlensing Events\\ in OGLE-IV data. I.~Events with Well-Separated Bumps}
\Author{
R.\,A.\,P.~~O~l~i~v~e~i~r~a$^1$,~~
R.~~P~o~l~e~s~k~i$^1$,~~
P.~~M~r~ó~z$^1$,~~
A.~~U~d~a~l~s~k~i$^1$,~~ \\
J.~~S~k~o~w~r~o~n$^1$,~~
M.~~M~r~ó~z$^1$,~~
M.\,K.~~S~z~y~m~a~ń~s~k~i$^1$,~~
I.~~S~o~s~z~y~ń~s~k~i$^1$,~~ \\
K.~~U~l~a~c~z~y~k$^{2,1}$,~~
P.~~P~i~e~t~r~u~k~o~w~i~c~z$^1$,~~
K.~~R~y~b~i~c~k~i$^{3,1}$,~~
P.~~I~w~a~n~e~k$^1$,~~\\
M.~~W~r~o~n~a$^{4,1}$
~~and~~
M.~~G~r~o~m~a~d~z~k~i$^1$~~
}
{$^1$ Astronomical Observatory, University of Warsaw, Al. Ujazdowskie 4,\\ 00-478 Warszawa, Poland \\
e-mail: raphael@astrouw.edu.pl \\
$^2$ Department of Physics, University of Warwick, Coventry CV4 7AL, UK \\
$^3$ Department of Particle Physics and Astrophysics, Weizmann Institute of Science, Rehovot 76100, Israel\\
$^4$ Department of Astrophysics and Planetary Sciences, Villanova University,\\ 800 Lancaster Avenue, Villanova, PA 19085, USA}

\Received{April 29, 2025}
\end{Titlepage}

\Abstract{Gravitational microlensing depends primarily on the lens mass and
  presents a larger occurrence rate in crowded regions, which makes it the
  best tool to uncover the initial mass function (IMF) of low-mass stars in
  the Galactic bulge. The bulge IMF can be obtained from the luminosity
  function measured with the Hubble Space Telescope if one knows
  the statistics of binary stellar systems in the bulge. We aim to analyze
  a statistically significant number of binary-lens/single-source and
  single-lens/binary-source events, in order to explore the lower-mass end
  of the bulge IMF even in unresolved binary systems. This paper deals with
  events with clearly separated bumps and no caustic crossing or approach,
  whereas other types will be analyzed in following works. A
  fully-automated approach in the search and modeling of binary events was
  implemented. Event detection was carried out with a modified version of
  the algorithm used in previous studies. Model fitting was carried out
  with Markov chain Monte Carlo and nested sampling methods, in order to
  find the most probable solution among binary lens or binary source
  models. We retrieved 107 binary events in Optical Gravitational Lensing
  Experiment (OGLE) light curves spanning ten years in 9 high-cadence and
  112 low-cadence fields toward the bulge. Several criteria were applied
  to reduce false positives, resulting in 59 most likely binary lenses and
  48 binary sources. The tools were effective in detecting a bona-fide sample
  of binary events, with a distribution of Einstein timescales around
  35--40~d and flat distributions for mass ratio and source flux
  ratio. After proper consideration of detection efficiency, the statistics
  for binary fraction and mass ratio will provide valuable constraints for
  the bulge IMF.}{gravitational lensing: micro -- Galaxy: bulge -- stars:
  mass function -- binaries: general}

\Section{Introduction}
\vskip14pt
The relative number of stars as a function of their mass, both at the
moment of birth (initial mass function, IMF) and currently (present-day
mass function, PDMF), is a fundamental ingredient in various fields of
astronomy.  Since the IMF affects observable properties of stellar
populations, including integrated properties of distant galaxies, several
studies have focused on determining whether the IMF slope is universal or
sensitive to environmental conditions and redshifts (see review by Bastian
\etal 2010). The IMF is well determined for the Solar neighborhood and
nearby star clusters (Duch\^ene and Kraus 2013), but there is still a great
effort to extend this knowledge for farther systems (\eg Da Rio \etal 2009,
Kalirai \etal 2013, Geha \etal 2013, Capellari \etal 2012). A top-heavy IMF
in the early Universe is often hypothesized to explain the excess of
bright, massive galaxies at $z\gtrsim7$, as recently observed with the
James Webb Space Telescope (\eg Steinhardt \etal 2023, Woodrum \etal 2024,
van Dokkum and Conroy 2024).

The Galactic bulge is dominated by a metal-rich, $\alpha$-enriched stellar
population of $\approx10$~Gyr, with formation timescale $<2$~Gyr (Clarkson
\etal 2008, Barbuy \etal 2018). This short timescale indicates that, for
stars below the main-sequence turn-off, the IMF is expected to be very
similar to the observed PDMF.  Measurements of the bulge PDMF for low-mass
stars are particularly hampered by the severe interstellar reddening, large
distances and, to a lower extent, crowding.

Precise measurements of the Galactic bulge luminosity function with the
Hubble Space Telescope have been confined to low-reddening windows:
Sagittarius Window ($\ell=1\zdot\arcd25$, $b=-2\zdot\arcd65$, Calamida
\etal 2015), Baade's Window ($\ell=1\zdot\arcd02$, $b=-3\zdot\arcd92$,
Holtzman \etal 1998), and a farther field in near-infrared ($\ell
=0\zdot\arcd28$, $b=-6\zdot\arcd17$, Zoccali \etal 2000). Calamida \etal
(2015) used proper motions to remove disk contamination, deriving the IMF
for the mass range 0.15--1.0~\MS\ as two power laws with a break at
0.56~\MS. The authors derived slightly different slopes for different binary
fractions (tested from 0\% to 100\%, but probably larger than 30\%) and
mass-ratio distributions, concluding that the highest uncertainty comes
from unresolved binaries, as pointed out by Bastian \etal (2010).

The masses of distant low-mass stars can be inferred from the analysis of
the Einstein timescale ($t_{\rm E}$) of a large sample of microlensing
events. The parameter $t_{\rm E}$ denotes the time needed for source and
lens to move by one Einstein ring radius ($\theta_{\rm E}$), which is
proportional to the square root of the lens mass.  The $t_{\rm E}$
distribution for a large sample of events was previously used to measure
the lens mass distribution and estimate their mass function by, \eg Han and
Gould (1996), Sumi \etal (2011), Wyrzykowski \etal (2015), Navarro \etal
(2017), and Mróz \etal (2019). The further use of galactic models allows
to estimate the IMF slope (Calchi Novati \etal 2008, Wegg \etal 2017, Mróz
\etal 2017), but its dependence with binarity statistics is usually
neglected or not measured directly. For example, Mróz \etal (2019)
estimated that 10\% of the events are binary or affected by higher-order
parameters, and accounted for them by rescaling the measured optical depths
and event rates.

It is estimated that most stars are located in multiple systems, with
smooth variations of the frequency and key properties along different
masses (Duch\^ene and Kraus 2013), however, the binarity of stars is still
poorly constrained in the Galactic bulge. The proper inclusion of multiple
systems in microlensing population studies can significantly affect the
inferred event rates and $t_{\rm E}$ distributions, as analyzed by Wegg
\etal (2017) and Abrams \etal (2025).  Apart from binary events with
obvious multiple peaks in the light curve, a considerable fraction of
binary lens (Di Stefano and Perna 1997) and binary source configurations
(Dominik 1998, Han and Jeong 1998) may present a single asymmetric peak and
be misclassified as non-binary events.

Here we search for binary microlensing events with separated bumps and no
caustic crossing or approach, in ten years of data from the fourth phase of
the Optical Gravitational Lensing Experiment (OGLE-IV, Udalski \etal 2015).
After analyzing other morphologies in this series of papers, we aim to
obtain the binary fraction and mass ratio from a statistically significant
sample of events.  These results will then be corrected by detection
efficiency and combined with the luminosity function from Calamida \etal
(2015) to constrain the bulge IMF.

Gravitational microlensing events occur when a massive object passes at a
close angle to a background star. If both stars are single systems and
considered point masses (point-source point-lens, hereafter PSPL), the lens
magnifies the observed flux following the Paczyński curve (Paczyński
1986):
$$A=\frac{u^2+2}{u\sqrt{u^2 + 4}},\eqno(1)$$
\noindent
where the separation $u$ depends on time as $u(t) = \sqrt{u_0^2 + \left (
    \frac{t-t_0}{t_{\rm E}} \right)^2}$. The parameter $t_0$ represents the
time of the closest approach, and the other two (impact parameter $u_0$ and
$t_{\rm E}$) are defined relative to $\theta_{\rm E}$, which is given by:
$$\theta_{\rm E}=\sqrt{\kappa M\left(\frac{\rm{au}}{D_l}-\frac{\rm{au}}{D_s} \right)},\eqno(2)$$
\noindent
where $\kappa=4G/(c^2{\rm au})=8.1\,{\rm mas}/\MS$ is a constant, $M$ is
the lens mass, and $D_l$ and $D_s$ are the lens and source distances,
respectively.

The fact that microlensing depends primarily on lens mass, not brightness,
allows to study objects in a wide mass range, including those undetectable
by current telescopes at the bulge distance (\eg M dwarfs, brown dwarfs,
free-floating planets -- Mróz \etal 2020). The event rate is noticeably
larger in dense fields, such as the Galactic bulge and disk, and depends on
the mass function and kinematics of the lenses. These facts combined make
microlensing a unique tool to study the bulge IMF down to low-mass stars,
including unresolved binary systems.

The existence of a binary companion with either the lens or source star
changes the light curve significantly. Binary source events (1L2S) are a
superposition of two PSPL events of same $t_{\rm E}$ if the source orbital
motion is ignored, with parameters: $t_{0,1}$, $u_{0,1}$, $t_{0,2}$,
$u_{0,2}$, and $t_{\rm E}$. It allows to measure the source flux ratio and
projected source separation relative to $\theta_{\rm E} D_s$. In order to
present two bumps and be distinguished from single source events, a 1L2S
event may not have too large source flux ratio or too small source
separations, which occur in 10--20\% of the events (Griest and Hu 1992,
Han and Jeong 1998).

On the other hand, calculating the magnification of binary lens events
(2L1S) requires solving a fifth-order complex polynomial with different
coefficients for each epoch, often showing complex caustic patterns (\eg
Mao and Paczyński 1991, Liebig \etal 2015). In this case, the mass ratio
$q$, the projected separation between the lens components $s$, and the
source trajectory angle $\alpha$ are added to the PSPL parameters. The
physical lens separation is obtained as $s \theta_{\rm E} D_l$. If there is
a caustic crossing or a close approach, the source angular size can be
further measured and modeling becomes more complex due to multiple
degeneracies. The analysis of events with caustics is addressed in
following papers of this series.

The OGLE-IV data (Udalski \etal 2015) contain more than 20\,000
microlensing events detected toward the Galactic bulge, alerted by the
Early Warning System (EWS -- Udalski 2003) since 2010.  Mróz \etal (2019)
considered data between 2010 and 2017, detecting 8000 events in the bulge
to produce the most accurate event rate and optical depth maps available to
date. The event rate depends on the mass function and kinematics of the
lenses, and is of the order of $10^{-6}\,\rm{yr}^{-1}$ and up to $\approx
250~{\rm deg}^{-2}\rm{yr}^{-1}$ for the densest OGLE fields. The optical
depth is defined as the probability that a source star is microlensed by a
foreground object, which depends on the mass density of stars and compact
objects (Kiraga and Paczyński 1994).  Both maps will be used in the last
steps of the present analysis.

This paper is structured as follows. The OGLE-IV data and a literature
compilation of binary events is presented in Section~2. Section~3 presents
the modifications applied to our event finding algorithm, in order to
detect binary events.  The model fitting pipeline applied to the detected
sample is described in Section~4. Section~5 presents the 107 detected
binary events, comparing the fitted binary source and binary lens models,
and followed by an initial population analysis toward IMF derivation.
Conclusions are drawn in Section~6.

\Section{Data: OGLE-IV and Benchmark Sample} The OGLE survey was created
after Paczyński (1986) proposed implementing long-term observations in
order to search for dark matter with microlensing. The first observations
were carried out in 1992 (Udalski \etal 1992) and the survey has collected
data nearly continuously since then, establishing as one of the largest
long-term photometric surveys. We used data from the OGLE-IV phase between
2010 June 29 ($\rm{JD} = 2455377$) and 2020 March 17 ($\rm{JD} = 2458926$,
due to pandemic shutdown).

The OGLE observations are obtained with the 1.3~m Warsaw Telescope, located
at the Las Campanas Observatory, Chile. The telescope is equipped with a
mosaic camera for OGLE-IV, containing 32 CCD detectors of $2048\times4102$
pixels, with a field of view of 1.4~deg$^2$ and a pixel scale of
$0\zdot\arcs26\,\rm{pixel}^{-1}$. The data reduction process is based on
the difference image analysis technique (DIA, Tomaney and Crooks 1996,
Alard and Lupton 1998, Woźniak 2000), in which a reference image of each
field is obtained from high-quality frames and subtracted from the
following ones to perform photometry. Udalski \etal (2015) describe the
instrument and data reduction pipeline in detail.

\begin{figure}[b]
\centerline{\includegraphics[width=9.5cm]{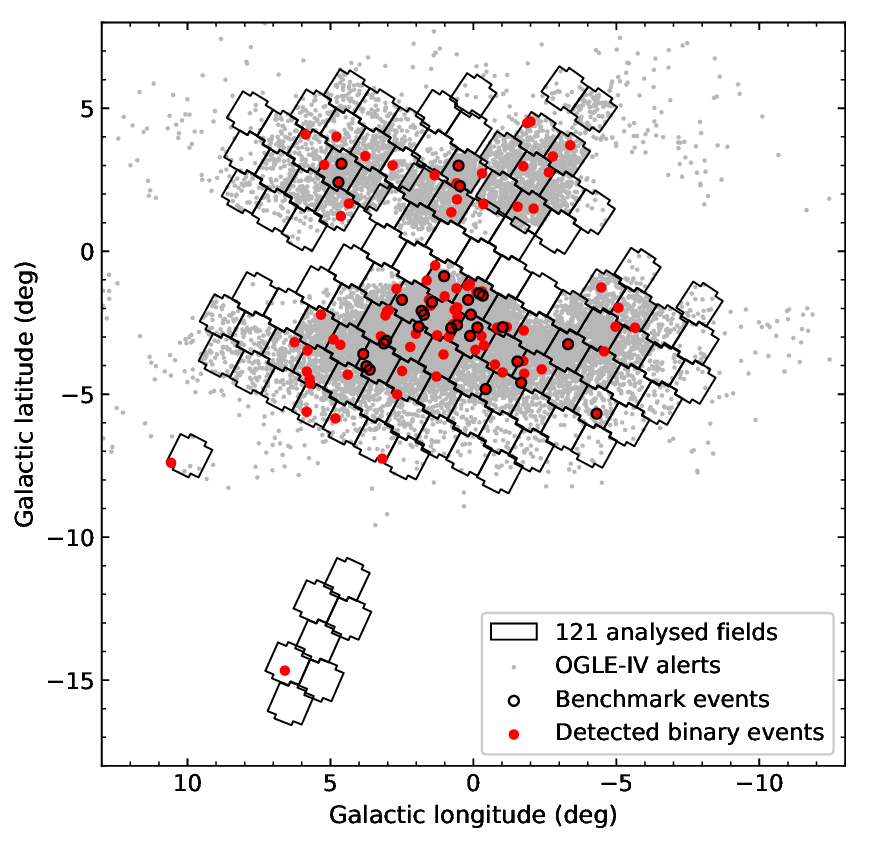}}
\FigCap{Distribution in galactic coordinates of the 121 analyzed OGLE
  fields, 20\,255 EWS alerts since 2010, 29 benchmark events and the
  detected 107 binary events. Around half of the binary candidates are
  toward high-cadence fields, therefore closer to the Galactic center.}
\end{figure}
The 121 OGLE fields analyzed in this work share different cadence values
along the 10-yr baseline. Nine fields have cadence of 20~min to 60~min
(10 to 30 datapoints per night), generating light curves of 4500--12\,000
datapoints (Mróz \etal 2017). The remaining fields have cadence ranging
from 60~min to less than one observation per night, with most of the
light curves containing between 100 and 2000 datapoints. We limit this
analysis to light curves with at least 120 datapoints. The coordinates and
cadence of all bulge fields can be retrieved in the OGLE database
website\footnote{\it https://ogle.astrouw.edu.pl/sky/ogle4-BLG/}, and
further details about the 121 analyzed fields are available in the Tables~6
and 7 from Mróz \etal (2019).

\MakeTableee{lccc}{12.5cm}{Sample of 29 benchmark binary microlensing events}
{\hline
\noalign{\vskip3pt}
OGLE ID      &  EWS ID  &  $t_{\rm bump, 1}$ & $t_{\rm bump, 2}$ \\
\noalign{\vskip3pt}
\hline
\noalign{\vskip3pt}
BLG500.22.46297  & OGLE-2013-BLG-0161 &  6409.9 & 7589.8 \\
BLG501.08.186097 & OGLE-2013-BLG-1010 & 6511.7 & 7135.7 \\
BLG501.13.138484 & OGLE-2011-BLG-0312 & 5684.9 & 8191.8 \\
BLG501.13.151818 & OGLE-2014-BLG-1367 & 6905.5 & 7589.5 \\
BLG501.20.76926  & OGLE-2018-BLG-0339 & 8339.6 & 8194.8 \\
BLG504.03.55458  & OGLE-2018-BLG-0190 & 8338.6 & 8152.9 \\
BLG504.03.189220 & --                 & 5426.5 & 5489.5 \\
BLG504.06.110803 & OGLE-2016-BLG-0594 & 7489.9 & 7617.6 \\
BLG504.28.163006 & OGLE-2019-BLG-1366 & 8721.5 & 8766.5 \\
BLG505.03.213693 & OGLE-2016-BLG-1013 & 7562.6 & 7937.5 \\
BLG505.05.14979  & OGLE-2017-BLG-0021 & 8206.8 & 7799.9 \\
BLG506.21.49242  & OGLE-2017-BLG-0107 & 7851.7 & 7803.9 \\
BLG506.23.129983 & OGLE-2017-BLG-1489 & 8642.9 & 8002.6 \\
BLG511.01.16650  & OGLE-2013-BLG-1515 & 6504.6 & 6583.6 \\
BLG511.01.152151 & OGLE-2014-BLG-0744 & 6792.9 & 7065.9 \\
BLG512.16.111826 & OGLE-2013-BLG-1173 & 6479.8 & 6525.6 \\
BLG534.09.159788 & OGLE-2011-BLG-0249 & 5667.7 & 5685.9 \\
BLG611.09.12112  & OGLE-2012-BLG-0128 & 6003.9 & 6544.5 \\
BLG611.21.139257 & OGLE-2015-BLG-2118 & 8524.9 & 7318.5 \\
\noalign{\vskip3pt}
\hline
\noalign{\vskip3pt}
BLG502.22.138016 & OGLE-2015-BLG-1406 & 7312.5 & 7198.7 \\
BLG507.04.99922  & OGLE-2015-BLG-0250 & 7263.5 & 7092.8 \\
BLG508.17.126114 & OGLE-2013-BLG-1045 & 6483.7 & 6469.5 \\
BLG510.11.45341  & OGLE-2012-BLG-0330 & 6033.8 & 6211.5 \\
BLG515.28.104226 & OGLE-2014-BLG-1156 & 7102.9 & 6855.7 \\
BLG519.19.16589  & OGLE-2015-BLG-0827 & 7140.7 & 7278.7 \\
BLG519.19.88012  & OGLE-2014-BLG-1454 & 6880.7 & 7102.8 \\
BLG519.29.136209 & OGLE-2015-BLG-0681 & 7249.7 & 7129.7 \\
BLG632.09.127659 & OGLE-2013-BLG-0702 & 6423.8 & 6485.5 \\
BLG632.22.132252 & OGLE-2019-BLG-0534 & 8592.8 & 8605.7 \\
\noalign{\vskip3pt}
\hline
\noalign{\vskip3pt}
\multicolumn{4}{p{8.5cm}}{
  The last two columns give the derived times of the separated bumps (HJD
  format, subtracted by 2450000). The horizontal line separates the events
  toward high- and low-cadence fields.}
}
Recently, the reference images were changed for several OGLE fields in the
central bulge and the photometry was reprocessed, including datapoints
after pandemic shutdown (August 2022, $\rm{JD}>2459800$). To ensure
consistency with previous works and the time range used in this work, we
used DIA photometry extracted using reference images constructed using
images from 2010--2012.

The event detection and model fitting (Sections~3 and 4) require a sample
of benchmark events with well-separated bumps, for which all major
implementations to the algorithms must prioritize their detection as binary
events. The thresholds were also optimized to detect this sample and
minimize the number of false positives. A sample of 29 benchmark events
was selected from visual inspection of an internal compilation of alerts
and literature events.

Fig.~1 shows the analyzed OGLE-IV fields, with the EWS alerts since 2010,
the benchmark sample and the detected sample of 107 binary events, all
naturally more concentrated in the most central fields. Table~1 presents
the sample of 29 benchmark events (19 in high-cadence and 10 in low-cadence
fields), for which the adopted algorithms were optimized and tested to
flag as binary events. The columns contain the OGLE identifier, the EWS
alert identifier, and the times of the two bumps. The more prominent bump
generated the alert, with the secondary bump affecting the provided PSPL
model only if it occurred in a earlier time.

\Section{Automated Detection of Binary Events} 
The binary microlensing events were searched in the same 121 OGLE-IV fields
toward the Galactic bulge as in Mróz \etal (2019), with a slightly longer
time range of 10~yr. More than 400 million sources were analyzed with a
modified version of the selection algorithm from Mróz \etal (2017, 2019),
now optimized to detect binary events with well-separated bumps and
subtract long-term linear trends in photometry.

\subsection{Description of the Original Algorithm}
The original algorithm considered a ``bump'' in the light curve as at least
three consecutive datapoints $3\sigma_{\rm base}$ above the baseline flux.
The baseline flux $F_{\rm base}$ and dispersion $\sigma_{\rm base}$ were
calculated using the datapoints outside of a 360-d window containing the
bump. Several windows were tested in steps of 80~d and removing $5\sigma$
outliers, and the best window was selected to minimize $\sigma_{\rm base}$.
If a bump was detected inside the window, the bump duration, amplitude,
significance ($\chi_{\rm bump}$, defined below), and peak time were
obtained comparing $F_{\rm base}$ and the points above $3\sigma_{\rm
base}$. A PSPL model was also fitted, in order to apply criteria based
on goodness-of-fit statistics at a later stage.

Datapoints outside of the bump window are expected to have constant flux\break
values to certain degree, in which large values of $\chi^2_{\rm base}/
\rm{dof}$ indicate that the event may present other type of variability.
The parameters $\chi^2_{\rm base} \rm{/dof}$ and $\chi_{\rm bump}$\break
have similar expressions but are calculated using different groups of epochs,
\ie the $N$ datapoints outside of the 360-d window and $N_{\rm bump}$
datapoints in the bump,\break respectively:
\setcounter{equation}{2}
\begin{align}
\chi^2_{\rm base}/\rm{dof} &= \frac{1}{N-1} \sum_{i=1}^N \frac{(F_i - F_{\rm base})^2}{\sigma_i^2} \\
\chi_{\rm bump} &= \sum_{j=1}^{N_{\rm bump}}{\frac{F_j - F_{\rm base}}{\sigma_j}} \,.
\end{align}

Mróz \etal (2019) adopted the following criteria to the light curves with
detected bumps: $\chi^2_{\rm base}/\rm{dof}\leq2.0$, $\chi_{\rm bump}> 32$,
and bump amplitude $A\geq0.10$. This guaranteed that other variability
types were discarded and that light curves with small or noisy bumps were
not included, as accounted for in the detection efficiency. At this point,
Mróz \etal (2017, 2019) reduced samples from hundreds of millions to
$\approx12\,000$ and $18\,000$, respectively. Further criteria based on the
goodness-of-fit of PSPL models led to final samples of 2617 and 5790
events, respectively.

Binary events were intentionally rejected in this algorithm when limiting
to a single bump or limiting goodness-of-fit statistics from a fitted PSPL
model, such as $\chi^2_{\rm fit}/\rm{dof}\leq2.0$. Moreover, events with
bumps separated by an interval larger than the window duration of 360~d were
also rejected, as occurred for eight benchmark events from Table~1. In such
cases, $\chi^2_{\rm base}/\rm{dof}$ may get higher than 2.0 as the
secondary bump is assumed to be part of the baseline, increasing its
dispersion and not selecting the event as a candidate. If a secondary bump
is allowed, checks for additional bumps may be implemented to discard other
types of variability.

\subsection{Implementations to Detect Binary Events}
In this work, in order to select clear binary events, we adopted stricter
thresholds compared to Mróz \etal (2017, 2019). For the high-cadence
fields, we searched for main bumps with at least seven consecutive
datapoints above $F_{\rm base}+3\sigma_{\rm base}$ and additional bumps
with at least five datapoints. For the low-cadence fields, we reduced these
minimum numbers to five and three datapoints, respectively, allowing to
detect shorter events.

The window length of 360~d and step of 80~d were preserved but an
additional window was used, in order to minimize $\chi^2_{\rm
base}/\rm{dof}$ specially in the cases with the two bumps largely
separated.  A PSPL model was fitted excluding the datapoints inside the
second window, but no cut in goodness-of-fit was applied to ensure that
light curves with asymmetries or multiple bumps are not rejected. The
threshold value considered to remove non-consecutive outliers was also
reduced from $5\sigma$ to $4\sigma$, in order to filter out noisy light
curves.

The adoption of two moving windows, slope correction and time binning
significantly increased computing time of the event finding code. All the
significant changes to the original algorithm are described and illustrated
in the following subsections. Last part of this section presents the tests
to obtain the best combination of thresholds for the binary events, in
order to reduce the number of false positives and maximize the detections
of genuine events.
\vskip3pt
{\it Slope correction to the Baseline Flux}
\vskip3pt
Microlensing light curves are expected to have a flat baseline out of the
bump duration, however, several factors can produce some variability.
Linear trends can be caused by proper motions, whereas intrinsic
variability can emerge from either the source star or the blended flux with
unlensed stars, and systematic noise may be produced by calibration issues,
detector sensitivity, or sky conditions.  Wyrzykowski \etal (2006) first
studied the baseline variability of the OGLE-III data, assuming
non-constant blending parameter, followed by Li \etal (2019) and Golovich
\etal (2022) who applied Gaussian process models to model the variability
for a single event and for many events in OGLE-III and OGLE-IV, respectively.

Golovich \etal (2022) presented the event BLG156.7.141434 (OGLE-III) as
example of an increase of 0.03~mag in the baseline level over seven years
and a less clear oscillatory behavior. They concluded that the baseline
variability is more pronounced in events with longer $t_{\rm E}$ and,
together with the omission of parallax effects, results in an
overestimation of the number of events with $t_{\rm E}>100$~d.

The slope of baseline photometry is smaller for our sample in general than
for BLG156.7.141434, with very few exceptions. Yet it is still important
to include a correction to account for events with bumps separated by a
long timescale.  In these cases, it is possible that the existence of the
binary star have introduced a bias on the determination of the source flux
in the reference images, generating a linear trend in the baseline.
Therefore, we apply a simpler correction compared to Golovich \etal (2022),
based on the correction of the slope of the baseline flux.

\begin{figure}[htb]
\centerline{\includegraphics[width=11.7cm]{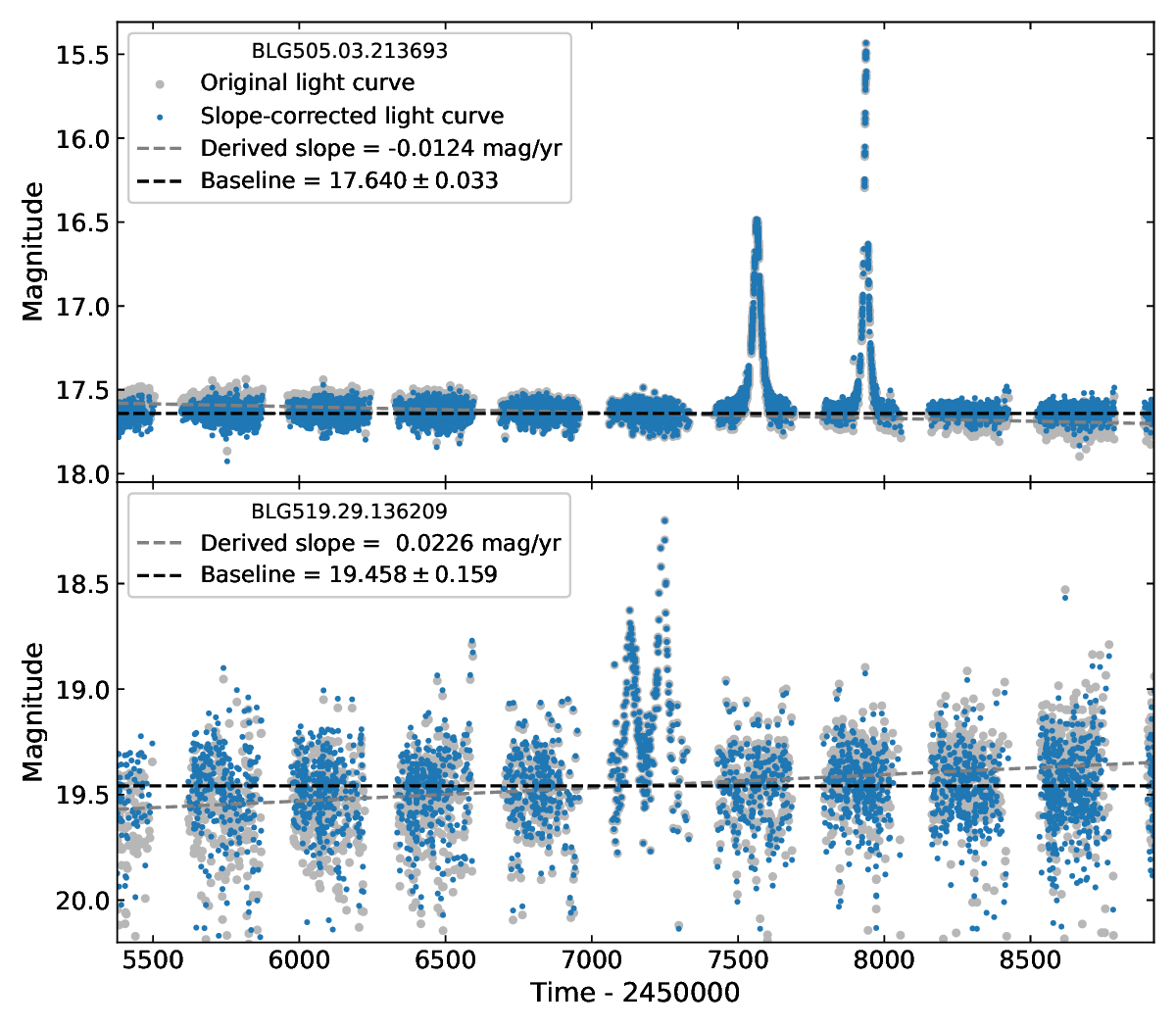}}
\FigCap{Light curve of the two benchmark events with largest slope
  correction, BLG505.03.213693 and BLG519.29.136209. Both would not be
  detected as binary in the uncorrected data.  The correction is more
  evident at the edges, as it fixed the middle point as the intersection
  point.}
\end{figure}

The choice of the two 360-d windows is carried out by moving all possible
combinations by 80~d. The pair that minimizes $\chi^2_{\rm base}/\rm{dof}$
is selected, strictly containing the two bumps in case of binary events.
Assuming that the baseline level varies linearly with time, the datapoints
outside of the windows are used to fit a straight line with fixed
intersection point as the middle time value ($t_{\rm middle}$) of the light
curve. The baseline flux $F_{\rm base}$ and $\sigma_{\rm base}$ are
computed with the flux values of all datapoints corrected using the derived
slope as:
$$F_{\rm new}=F_{\rm orig}-{\rm slope}\times(t-t_{\rm{middle}}).\eqno(5)$$

This correction reduces the baseline flux uncertainty ($\sigma_{\rm base}$)
and $\chi^2_{\rm base}/{\rm dof}$, allowing the detection of some PSPL and
binary events previously undetected. The fact that the flux-corrected data
follow a more constant $F_{\rm base}$ value also ensures that the correct
non-consecutive points are removed as outliers.  The derived slopes of the
analyzed binary sample are provided in the tables in Section~5, with an
average value of 0.0014\,mag/yr and maximum value of $\approx0.02$~mag/yr.

Fig.~2 presents the two benchmark events with the largest absolute values
derived for the slope: BLG505.03.213693 ($- 0.0124$\,mag/yr) and
BLG519.29.136209 ($0.0226$\,mag/yr). Both light curves have a clear
difference in the baseline at the edges, and would not be detected as
binaries without the slope correction: $\chi^2_{\rm base}/{\rm dof}$
reduced from 2.049 to 1.155 and from 1.579 to 1.384, respectively. The
values of $\sigma_{\rm base}$ were also reduced from 0.057 to 0.043 and
from 0.040 to 0.038, respectively, which helped to obtain the correct
number of datapoints above $3\sigma_{\rm base}$ and more precise values for
$\chi_{\rm bump}$ and duration.
\vskip3pt
{\it Bump Duration and Check for Additional Bumps}
\vskip3pt
Binary microlensing events can present several light curve morphologies,
specially with caustic patterns. In the case of well-separated bumps, the
light curve must return to the level $F_{\rm base}+3\sigma_{\rm base}$
around the main bump before rising into the secondary bump, in order to be
detected.  The duration of a detected bump is computed as the time range
where the datapoints have flux values above $F_{\rm base}+3\sigma_{\rm
base}$.

In events with lower amplitude, however, the datapoints appear more
dispersed around the bumps and the duration can be underestimated as a
single datapoint with flux below the thresholds terminates the duration
evaluation. Since a few benchmark events had short durations and no second
bumps detected, the bump duration calculation was implemented using the
median flux in time bins of 1~d (or 3~d for low-cadence fields). Using
this approach, the bump points are only considered to be above the
threshold level if the median flux of the datapoints in each bin is above
it on both sides.

\vskip3pt
The routine for bump finding is applied iteratively up to three times:
first to the original datapoints, then to datapoints with the main bump
removed and, if a secondary bump is found, it is applied to the datapoints
removing this bump. The bump removals are carried out using the derived
bump duration and checking if the median flux of the bin is still above
$3\sigma_{\rm base}$. When the median gets below this threshold, the bump
removal is stopped. Additional durations and thresholds are tested, in
order to avoid that residual datapoints are left in the bump wings and make
sure it will not mark a false bump.

\begin{figure}[htb]
\centerline{\includegraphics[width=12cm]{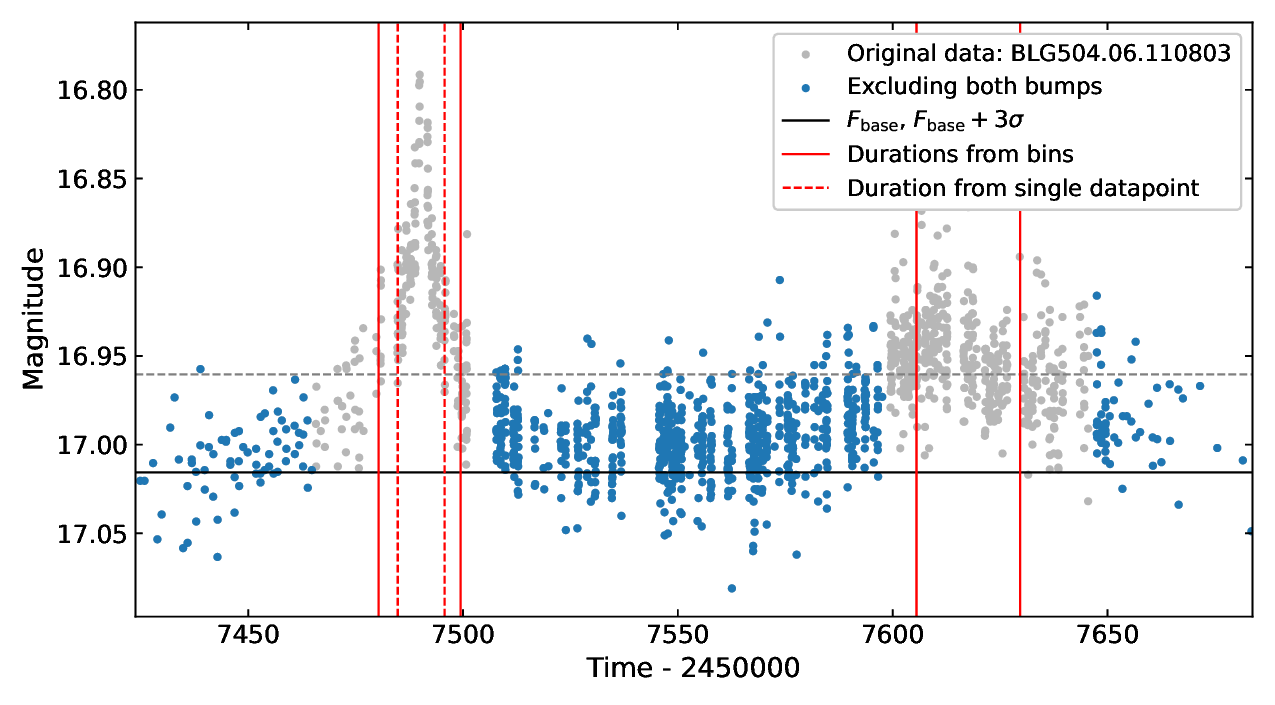}}
\FigCap{Light curve around the two bumps detected for BLG504.06.110803.
  The $3\sigma_{\rm base}$ level above the baseline shows that using
  individual points to derive the duration would underestimate the duration
  of the main bump ($t\approx7490$) and not find enough consecutive points
  in the other ($t\approx7618$).}
\end{figure}
\vskip3pt
The time binning was very effective to detect some benchmark events and
remove bumps in their light curves, \eg for BLG501.08.186097,
BLG501.31.52436, and BLG504.06.110803. Fig.~3 shows the light curve of
BLG504.06.110803, where the main bump had underestimated duration and the
secondary one was not detected using the algorithm without binning. The
new approach allowed to derive a more realistic duration for the first bump
(from 10.9~d to 19.1~d) and detect the other bump with duration of 24.1~d.
The figure also shows that the removal of the two bumps extend beyond their
duration (gray compared to blue points), to ensure that the bump wings are
also removed.

\vskip3pt 
If the event is identified to have a secondary bump but not a
third one, it is flagged as binary. The event is then tested into several
thresholds for the parameters of both bumps (as described in the next
subsection) and the model fitting routine is applied with binary source and
binary lens models (Section~4).

\newpage
{\it Different Thresholds for High- and Low-Cadence Fields}
\vskip3pt
The selection algorithm provides parameters for the baseline and the two
detected bumps: $\chi^2_{\rm base}/\rm{dof}$ of the baseline points around
the mean flux, as well as number of points, $\chi_{\rm bump}$, amplitude,
and duration for both bumps.  In addition to the fitted free parameters,
the model fitting also provides parameters that indicate the quality of the
fit, considering all datapoints or those around the bump. Studies of large
microlensing samples usually apply a series of selection criteria to reduce
the sample of detected candidates and filter out the false positives (\eg
Alcock \etal 2000, Wyrzykowski \etal 2009, Sumi \etal 2011, Wyrzykowski
\etal 2015, Mróz \etal 2017, 2019).

In the present work, different sets of thresholds were applied to high- and
low-cadence fields. For both cases, the thresholds were selected to
optimize the detection of benchmark events in testing fields and find the
best compromise between the size and purity of the sample, rejecting the
largest number of false positives without removing too many genuine binary
events. The applied thresholds are slightly stricter than in papers by
Mróz \etal (2017, 2019), which were focused in a homogeneous sample of PSPL
events.  Detection efficiency corrections to be applied in future papers
will account for genuine events skipped at this stage.

For high-cadence fields, the three fields with the largest number of EWS
alerts and binary events (internal sample) were used as testing fields:
BLG505, BLG501, and BLG504. Initially, a set of looser thresholds was
applied to $\chi^2_{\rm base}/\rm{dof}$ and parameters of the first bump to
select a few hundred candidates in these fields. After visual inspection,
a sample of ``genuine'' events with two clear, well-separated bumps was
selected. Cumulative distribution functions were drawn for $\chi^2_{\rm
  base}/\rm{dof}$, significance, amplitude, and duration of both bumps,
showing separately the distribution of PSPL and binary detections, as well
as the genuine and benchmark events. Several threshold values were tested,
in order to not remove any benchmark events and select the largest sample
of genuine events. The final adopted values and number of remaining
candidates are given in Table~2.

\MakeTable{llc}{12.5cm}{Selection criteria for the nine high-cadence OGLE-IV fields}
{\hline
\noalign{\vskip4pt}
Criteria      &  Remarks  &  Number \\
\noalign{\vskip4pt}
\hline
\noalign{\vskip4pt}
All stars  & -- &
49\,951\,279 \\
\noalign{\vskip4pt}
\hline
\noalign{\vskip4pt}
& {\it Main bump:} \\
$n_{\rm stars}\geq7$                & Consecutive points above $F_{\rm base}+3\sigma_{\rm base}$ \\
$\chi^2_{\rm base}\rm{/dof}\leq1.5$ & No variability in the baseline datapoints \\
$\chi_{\rm bump}>100$               & Significance of the bump compared to the baseline \\
$A>0.15$~mag                        & Rejecting low-amplitude variables &7\,316 \\
\noalign{\vskip4pt}
\hline
\noalign{\vskip4pt}
& {\it Secondary bump:} \\
$n_{\rm bump}=2$         & Rejecting events with one or more than two bumps \\
$\chi_{\rm bump,\,2}>60$ & Significance of the secondary bump \\
$A_2>0.1$~mag            & Rejecting low-amplitude secondary bumps \\
duration$_1 \geq 5.0$\,d & Avoid short duration in main bump \\
duration$_2 \geq 2.0$\,d & Avoid secondary bumps with very short duration &
121 \\ 
\noalign{\vskip4pt}
\hline
\noalign{\vskip4pt}
& {\it After the fitting:} \\
$\chi^2_{\rm fit}\rm{/dof}\leq2.0$    & Quality of the fit, using all datapoints \\
$F_{\rm s} > 0$                       & Rejecting fits with negative source flux \\
$|t_{0,1}-t_{0,2}|>9$~d               & Removing too close bumps, not well-separated \\
$0.01<\max(u_{0,1},u_{0,2})<1.6$      & Constrain amplitude for main bump (reject caustics) \\
$t_{\rm E}< 200$~d                    & Maximum timescale (reject caustics) \\
$s_{\rm 1L2S}<120$, $s_{\rm 2L1S}<250$& Removing large separations, probable CVs & 53 \\
\noalign{\vskip4pt}
\hline
\noalign{\vskip4pt}
\multicolumn{3}{l}{The criteria regarding the main bump are comparable to those from Mróz \etal (2017).}\\
}

The low-cadence fields, on the other hand, contain a much smaller number of
epochs (median of $\approx500$ epochs) and consequently less known events per
field. For this reason, a different approach for assigning thresholds
values was adopted, based on degrading high-cadence data. Data for the 29
benchmark events were degraded by iteratively selecting a random datapoint
and removing all other datapoints from the same night until the number of
500 epochs is reached. The resulting light curve has approximately one
datapoint each few nights, closely resembling the observations in the
low-cadence fields. The event finder algorithm was applied to 10 random
realizations of each benchmark event, showing the limitations of the data
in each threshold. For example, we noted that events shorter than 5~d
would hardly be detected in data with 500 epochs, whereas the $\chi^2_{\rm
  base}/\rm{dof}$ distribution does not change significantly compared to
the high-cadence fields.

\newpage
With the events detected in the degraded data replacing the sample of
genuine events in high-cadence data, similar cumulative distribution
functions were drawn for a selection of nine low-cadence fields.  The
selection of best thresholds was carried out ensuring the detection of all
benchmark events, maximizing the detection of degraded events, and
minimizing the false positives count. The final adopted values in
low-cadence fields and number of candidates are given in Table~3.

\MakeTable{llc}{12.5cm}{Selection criteria for the 112 low-cadence OGLE-IV fields}
{\hline
\noalign{\smallskip}
Criteria      &  Remarks  &  Number \\
\noalign{\smallskip}
\hline
\noalign{\smallskip}
All stars  & -- &
353\,813\,973 \\
\noalign{\smallskip}
\hline
\noalign{\smallskip}
& \textit{Main bump:} \\
$n_{\rm stars} \geq 5$  & Consecutive points above $F_{\rm base}+3\sigma_{\rm base}$ \\
$\chi^2_{\rm base}\rm{/ dof}\leq1.4$
& No variability in the baseline datapoints \\
$\chi_{\rm bump}>50$ & Significance of the bump compared to the baseline \\
$A>0.12$~mag & Rejecting low-amplitude variables &
63\,487 \\
\noalign{\smallskip}
\hline
\noalign{\smallskip}
& \textit{Secondary bump:} \\
$n_{\rm bump}=2$ & Rejecting events with one or more than two bumps \\
$\chi_{\rm bump,2}>20$ & Significance of the secondary bump \\
$A_2 >0.05$~mag & Rejecting low-amplitude secondary bumps \\
duration$_1\geq 5.0$~d & Avoid short duration in main bump \\
duration$_2\geq 3.0$~d & Avoid secondary bumps with very short duration & 342 \\
\noalign{\smallskip}
\hline
\noalign{\smallskip}
& \textit{After the fitting:} \\
$\chi^2_{\rm fit}\rm{/dof}\leq1.5$ & Quality of the fit, using all datapoints \\
$F_{\rm s}>0$ & Rejecting fits with negative source flux \\
$|t_{0,1}-t_{0,2}|>9$~d & Removing too close bumps, not well-separated \\
$0.01<\max(u_{0,1},u_{0,2})<1.6$ & Constrain amplitude for main bump (reject caustics) \\
$t_{\rm E}<120$~d & Maximum timescale (reject caustics) \\
$s_{\rm 1L2S}<70$, $s_{\rm 2L1S}<70$ & Removing large separations, probable CVs & 66 \\
\noalign{\smallskip}
\hline
\noalign{\smallskip}
\multicolumn{3}{p{12cm}}{It can be noted that most of the criteria are looser
 than for high-cadence fields.}\\
}

The last lines of Tables~2 and 3 present the thresholds applied after the
fitting, rejecting fits with the negative source flux ($F_{\rm s}$) and
limiting the derived parameters.  Limits in $u_{0,i}$ and upper limit in
$t_{\rm E}$ were applied to avoid events with caustic patterns. The last
criteria is based on the separation from 1L2S and 2L1S models, with the
former as $s_{\rm 1L2S}=\sqrt{\tau^2+\max(u_{0,1},u_{0,2})}$, where
$\tau=(t_{0,1}-t_{0,2})/t_{\rm E, 1L2S}$. The exact value for this
threshold was obtained by comparing the binary candidates cross-matched
with the catalog of cataclysmic variables (CVs) given in Mróz \etal
(2015). Cataclysmic variables such as dwarf novae can present light curves
with short, bright bumps similar to microlensing, but with brightening
steeper than the fading.  In this work, the ones with short recurrence
times were probably not detected for having more than two bumps, but the
ones with longer recurrence times can be incorrectly detected as binaries
with large separations.

\MakeTable{lcccl}{12.5cm}{Sample of 12 false positives detected after visual inspection}
{\hline
\noalign{\smallskip}
Event      &  R.A.  &  Dec.  &
EWS ID & Comment \\
\noalign{\smallskip}
\hline
\noalign{\smallskip}
BLG501.17.60898  & 17:54:56.57 & $-$29:48:20.1 & -- & Multiple bumps \\
BLG502.31.116902 & 17:50:02.61 & $-$32:55:45.2 & -- & Probable cataclysmic variable \\
BLG504.27.87336  & 17:59:03.94 & $-$27:28:03.4 & OB120011 & Unrelated PSPL neighbor \\
BLG504.27.88595  & 17:59:03.87 & $-$27:28:03.2 & OB180898 & Unrelated PSPL neighbor \\
BLG506.23.129457 & 17:55:50.71 & $-$30:25:06.2 & OB190802 & Ghost of BLG506.23.129983 \\
BLG511.21.2207   & 18:03:12.67 & $-$27:19:27.5 & -- & Second bump is a false positive \\
BLG513.16.76279  & 18:00:10.40 & $-$30:04:31.7 & -- & Second bump is a false positive \\
BLG523.23.116646 & 18:12:31.65 & $-$27:16:37.1 & -- & Ghost of BLG523.23.114690 \\
BLG530.01.107532 & 18:21:22.27 & $-$27:18:54.8 & -- & Probable cataclysmic variable \\
BLG617.25.73581  & 17:12:59.93 & $-$29:30:33.6 & -- & Probable cataclysmic variable \\
BLG652.24.54795  & 17:39:03.73 & $-$26:01:33.0 & OB170428 & Second bump is a false positive \\
BLG717.25.25730  & 18:05:42.73 & $-$22:51:19.5 & -- & Probable Mira variable \\
\noalign{\smallskip}
\hline
}
A visual inspection verified that a small sample of 12 candidates are clear
false positives, as listed in Table~4. They consist of light curves with
either multiple bumps or no second bump, are more probably cataclysmic or
Mira variables, or occur on nearby stars. After removing these false
positives, the final sample contains 107 binary events, 48 in the
high-cadence fields and 59 in the low-cadence fields. With very few
genuine events removed, we conclude that the automated pipeline was
effective in detecting the benchmark events, and that the thresholds did a
good compromise between the size and purity of the sample.

\Section{Model Fitting with {\sf MulensModel}}
Microlensing light curves with two widely separated bumps can be closely
reproduced by both binary-source models and binary-lens models, making it
hard to distinguish which one best reproduces the data. For this reason,
we applied two sampling methods to the candidate events: Markov chain Monte
Carlo (MCMC) with {\sf emcee} (Foreman-Mackey \etal 2013) and nested
sampling with {\sf UltraNest} (Buchner 2021).

The {\sf MulensModel} package (Poleski and Yee 2019) was used to model the
binary candidates. {\sf MulensModel} provides several methods for
calculating the magnification, allowing to simulate the models of interest
and obtain the $\chi^2$ that compares the observed light curves to the
models. For binary source models, the effective magnification is simply
given by the weighted sum of the magnifications of the two sources.  For
binary lens models with point source, the fifth-order complex polynomial is
solved using the Skowron and Gould (2012) root solver implemented by Bozza
\etal (2018).

An initial first step for a fitting pipeline is to fit 1L2S models, which
is easily fitted in {\sf MulensModel} with no clear degeneracies. However,
in order to derive more precise 1L2S parameters and get reasonable starting
parameters for the 2L1S models, we implemented a separation of the data in
two parts to fit two PSPL models independently.  The epoch that divides the
data into two is obtained as the minimum flux between the two bumps of an
initial 1L2S fit. The PSPL fitting is carried out for each segment of the
data, fixing the blending flux ($F_{\rm bl}$) to 0 and iteratively
subtracting from the datapoints of one segment the PSPL model fitted to the
other segment.

Fig.~4 illustrates the results obtained with this procedure to the binary
event BLG511.14.135138, with the epoch of minimum flux above the baseline.
The data subtraction in the region between the bumps is particularly
important in such events with mixed bumps, which will be evaluated in
future papers.

\begin{figure}[htb]
\centerline{\includegraphics[width=11.7cm]{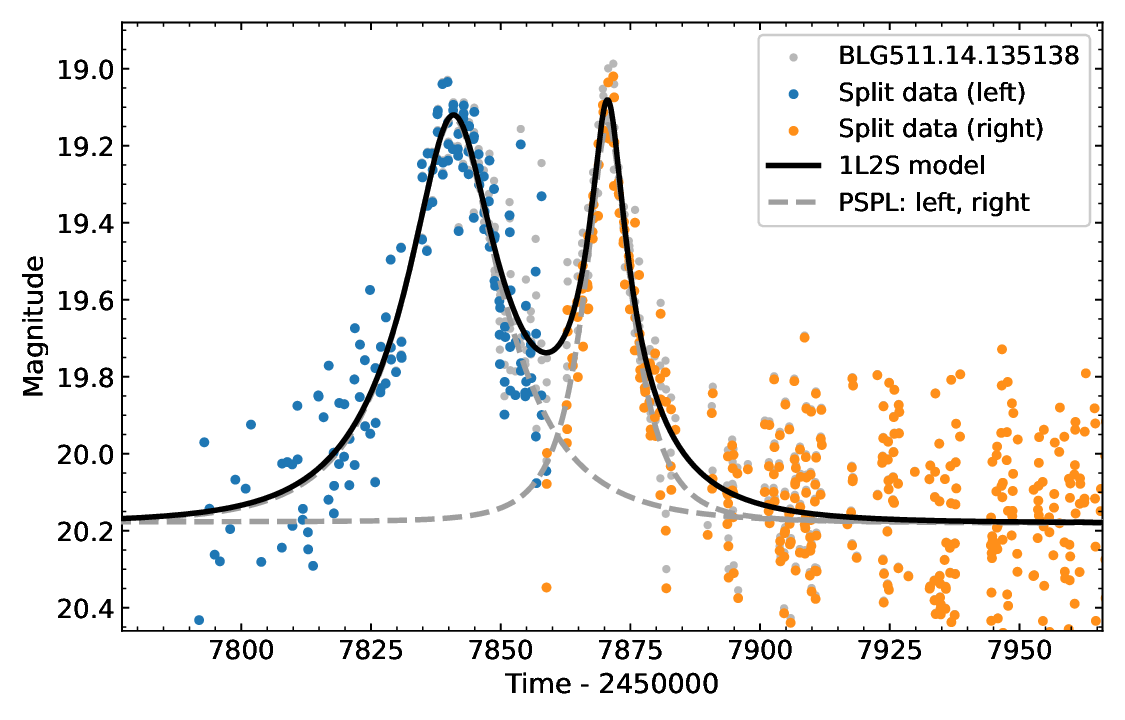}}
\FigCap{ Example of data separation for BLG511.14.135138, using the
  datapoint with minimum flux between the two bumps as the limit. Two PSPL
  fits are obtained independently, obtaining good initial parameters for
  the 1L2S and 2L1S fittings.}
\end{figure}

Following the conventions by Skowron \etal (2011) and Jung \etal (2015), we
explored the representation of the lens plane with the source trajectory
relative to it to get the 2L1S parameters in terms of the parameters of two
PSPL models. Starting from the equations for mass ratio and $\theta_{\rm
E}$, and using the geometry of the system, we derive:
\setcounter{equation}{5}
\vspace*{-4pt}
\begin{align}
q         &=\frac{t_{\rm E,2}^2}{t_{\rm E,1}^2}\\
t_0       &=\frac{q\times t_{0,2}+t_{0,1}}{1+q}\\
u_0       &=-\frac{u_{0,1}\mp q\times u_{0,2}}{1+q}\\
t_{\rm E} &=\sqrt{t_{\rm E,1}^2 + t_{\rm E,2}^2} \\
s         &=\frac{s^\prime + \sqrt{{s^\prime}^2 + 4}}{2} \qquad s^\prime = \sqrt{\frac{(t_{0,2} - t_{0,1})^2}{t_{\rm E,1}^2 + t_{\rm E,2}^2} + (u_{0,1} \pm u_{0,2})^2} \\
\alpha    &=\arctan \left ( \frac{u_{0,1} \pm u_{0,2}}{t_{0,2} - t_{0,1}} \times \sqrt{t_{\rm E,1}^2 + t_{\rm E,2}^2} \right )\,.
\end{align}
In this set of equations for the initial parameters, the plus-minus signs
indicate the two possible configurations of the system, with the source
trajectory passing between or beyond the two lenses, respectively.  Both
solutions are fitted in the current pipeline and the one with smaller
$\chi^2_{\rm fit}$ is considered.

Two prior distributions are applied to constrain $t_{\rm E}$ and blending
flux values. A Gaussian prior in $t_{\rm E}$ follows the distribution
derived in Mróz \etal (2020), which peaks around 30~d. The other prior is
a Gaussian distribution that smoothly disfavors too negative blending flux
values. The choice of the preferred model among 1L2S and 2L1S is primarily
based on the $\chi^2_{\rm fit}/{\rm dof}$ values. Since some events
present very close values for these models, if the difference in
$\chi^2_{\rm fit}/{\rm dof}$ is smaller than 0.001, we prioritize the model
with the smaller Bayesian evidence value (derived with {\sf UltraNest}),
smaller amount of negative blending flux and derived $t_{\rm E}$ closer to
the peak of the prior.

The MCMC fitting was carried out with a number of walkers and steps large
enough to allow a clear convergence of the chains and efficient sampling of
the posterior distribution in all the parameters. The {\sf UltraNest}
fitting uses the MLFriends algorithm (Buchner 2016, 2019) to allow a
minimum number of live points to explore the parameter space and evaluate
the evidence. Since {\sf UltraNest} was less efficient than MCMC, we
limited the parameter space to $5\sigma$ around the MCMC solution for
binary source and $3\sigma$ around the binary lens solution.

\Section{Results and Discussion}
The analysis of the high- and low-cadence fields resulted in 48 and 59
binary events, respectively, after applying all the detection and fitting
criteria.

Table~5 presents information for the 48 events detected in the high-cadence
fields, namely the coordinates, the derived slope and baseline magnitude
values (see Section~3), the preferred model (see Section~4), and the alert
identifiers. The baseline magnitude is given as $I_{\rm bl}$, without
correcting by blending because 1L2S and 2L1S models have different numbers of
source fluxes. Table~6 presents the same information for the 59 events
detected in the low-cadence fields.

In the high-cadence OGLE fields, only eight events were not alerted by
OGLE-EWS, MOA (Microlensing Observations in Astrophysics, Bond \etal 2001),
or KMTNet (Korean Microlensing Telescope Network, Kim \etal 2016). In the
low-cadence, instead, a larger number of 18 were not alerted, indicating that
the thresholds were looser. Most of the skipped events contain low-amplitude
bumps or a noisier baseline level, hampering the detection by the alert
systems.

Figs.~5 and 6 present a comparison between 1L2S and 2L1S fits obtained for
one event in low-cadence and one in high-cadence fields: BLG615.27.39817 and
BLG505.03.213693, respectively. The difference between the models is better
shown in the residual and cumulative $\Delta \chi^2=\chi^2_{\rm fit,
2L1S}-\chi^2_{\rm fit, 1L2S}$ plots, specially around the bumps. For
BLG615.27.39817, the difference in $\chi^2_{\rm fit}/\rm{dof}$ is slightly
larger than 0.001, but the derived values for evidence, blending flux, and
$t_{\rm E}$ also favor the 1L2S model. The corner plots of the preferred
models present the correlations between the parameters, as well as the
marginalized posterior distribution in the diagonal panels.

\newpage
\begin{landscape}
\renewcommand{\arraystretch}{1.3}
\renewcommand{\tabcolsep}{9pt}
\MakeTableee{lccrcccc}{19cm}{Sample of 48 events detected in the nine high-cadence fields}
{\hline
\noalign{\smallskip}
Event & R.A.      & Dec.    & \multicolumn{1}{c}{Slope} & $I_{\rm bl}$ & Pref. & EWS ID & Other IDs \\
      & [h:m:s]   & [d:m:s] & [mag/yr]                  &  [mag]       &       &        &  \\
\noalign{\smallskip}
\hline
\noalign{\smallskip}
BLG500.04.66459              &17:52:07.04&$-29{:}05{:}48.4$&$-0.0091$&$20.150\pm0.228$&2L1S&OB180508&KB182408\\
BLG500.09.56443              &17:54:09.62&$-28{:}53{:}00.7$&$ 0.0010$&$18.712\pm0.069$&1L2S&OB121195&\\
BLG500.22.46297$^{\dagger}$  &17:51:26.46&$-28{:}30{:}36.7$&$-0.0033$&$17.489\pm0.020$&1L2S&OB130161&MB13233, KB160449\\
BLG500.27.59448              &17:53:26.15&$-28{:}03{:}43.6$&$-0.0017$&$20.240\pm0.222$&1L2S&&\\
BLG500.31.22754              &17:50:42.08&$-28{:}03{:}01.9$&$-0.0003$&$18.507\pm0.042$&2L1S&&\\
BLG501.08.186097$^{\dagger}$ &17:54:34.64&$-29{:}59{:}49.4$&$-0.0015$&$16.533\pm0.012$&2L1S&OB131010&MB13514\\
BLG501.13.138484$^{\dagger}$ &17:50:57.19&$-30{:}01{:}07.5$&$-0.0015$&$19.210\pm0.093$&1L2S&OB110312&\\
BLG501.13.151818$^{\dagger}$ &17:50:58.34&$-29{:}52{:}23.4$&$-0.0009$&$18.441\pm0.049$&2L1S&OB141367&KB161999\\
BLG501.14.93030              &17:50:25.56&$-29{:}55{:}06.0$&$ 0.0070$&$20.258\pm0.243$&1L2S&OB151355&\\
BLG501.20.76926$^{\dagger}$  &17:52:44.65&$-29{:}39{:}16.1$&$-0.0023$&$16.748\pm0.014$&2L1S&OB180339&MB18280, KB182211\\
BLG501.22.125510             &17:50:58.25&$-29{:}45{:}54.2$&$ 0.0003$&$20.579\pm0.292$&1L2S&OB180386&\\
BLG501.30.90272              &17:50:52.84&$-29{:}22{:}59.9$&$-0.0047$&$19.471\pm0.114$&1L2S&OB131031&\\
BLG501.31.52436              &17:50:24.78&$-29{:}25{:}08.0$&$-0.0025$&$17.507\pm0.021$&1L2S&OB111029&\\
BLG504.03.55458$^{\dagger}$  &17:58:21.96&$-28{:}34{:}36.2$&$ 0.0005$&$18.411\pm0.052$&2L1S&OB180190&KB181115\\
BLG504.03.189220$^{\dagger}$ &17:58:01.10&$-28{:}26{:}37.0$&$-0.0009$&$16.182\pm0.008$&1L2S&&\\
\noalign{\smallskip}
\hline
\noalign{\smallskip}
\multicolumn{8}{p{19.1cm}}{Table contains first 15 objects, full list of
objects is available in the {\it Acta Astronomica Archive:} {\it
https://acta.astrouw.edu.pl/acta/2025/oli\_75\_1/table5.txt}. The columns
provide the OGLE ID, derived slope, total {\it I}-band magnitude of the
baseline, the preferred model, and the IDs from alert systems. Alerts
starting with MB and KB come from the MOA and KMTNet surveys. Events with
$^\dagger$ are present in the benchmark sample from Table~1.}}
\end{landscape}
\newpage
\newpage
\begin{landscape}
\renewcommand{\arraystretch}{1.3}
\renewcommand{\tabcolsep}{11pt}
\MakeTableee{lccrcccc}{19cm}{Sample of 59 events detected in the nine low-cadence fields}
{\hline
\noalign{\smallskip}
Event & R.A.    & Dec.    & \multicolumn{1}{c}{Slope} & $I_{\rm bl}$ & Pref. & EWS ID & Other IDs \\
      & [h:m:s] & [d:m:s] & [mag/yr]                  &  [mag]       &       &        &  \\
\noalign{\smallskip}
\hline
\noalign{\smallskip}
BLG502.22.138016$^{\dagger}$&17:50:40.21&$-33{:}26{:}52.2$&$-0.0049$&$19.742\pm0.135$& 2L1S & OB151406 & \\
BLG503.25.47961             &17:48:36.20&$-34{:}39{:}20.4$&$ 0.0006$&$18.324\pm0.040$& 2L1S & OB140350 & MB14115\\
BLG507.04.99922$^{\dagger}$ &17:57:23.48&$-32{:}13{:}35.6$&$ 0.0054$&$19.554\pm0.118$& 2L1S & OB150250 & \\
BLG507.08.126798            &18:00:10.70&$-31{:}57{:}58.7$&$-0.0039$&$20.045\pm0.221$& 1L2S & OB191153 & KB190724\\
BLG507.18.91970             &17:59:39.94&$-31{:}35{:}53.6$&$-0.0058$&$20.199\pm0.206$& 2L1S &  & \\
BLG508.13.123210            &17:56:31.31&$-33{:}05{:}48.9$&$ 0.0020$&$18.136\pm0.034$& 2L1S &  & \\
BLG508.17.126114$^{\dagger}$&18:00:08.38&$-32{:}42{:}21.6$&$-0.0025$&$18.095\pm0.040$& 2L1S & OB131045 & \\
BLG508.20.29308             &17:58:34.61&$-32{:}38{:}31.9$&$ 0.0014$&$18.995\pm0.090$& 1L2S & OB131046 & \\
BLG508.30.59690             &17:56:49.34&$-32{:}23{:}46.7$&$ 0.0008$&$18.526\pm0.041$& 2L1S &  & \\
BLG510.11.45341$^{\dagger}$ &17:58:26.32&$-35{:}31{:}41.6$&$ 0.0001$&$18.493\pm0.032$& 1L2S & OB120330 & \\
BLG513.08.44056             &18:06:01.28&$-30{:}01{:}09.4$&$ 0.0012$&$19.930\pm0.206$& 1L2S &  & \\
BLG513.13.113120            &18:02:20.38&$-29{:}51{:}40.7$&$-0.0008$&$17.707\pm0.027$& 2L1S &  & \\
BLG515.28.104226$^{\dagger}$&18:03:55.15&$-31{:}44{:}13.8$&$ 0.0002$&$17.122\pm0.016$& 2L1S & OB141156 & MB14399\\
BLG518.20.105927            &18:08:57.85&$-26{:}33{:}07.6$&$-0.0046$&$19.121\pm0.085$& 1L2S & OB180599 & KB193216\\
BLG518.29.27013             &18:08:44.20&$-26{:}15{:}01.9$&$-0.0064$&$18.766\pm0.062$& 2L1S & OB190279 & \\
\noalign{\smallskip}
\hline
\multicolumn{8}{p{19.1cm}}{Table contains first 15 objects, full list of
objects is available in the {\it Acta Astronomica Archive:} {\it
https://acta.astrouw.edu.pl/acta/2025/oli\_75\_1/table6.txt}. The columns
provide the OGLE ID, derived slope, total {\it I}-band magnitude of the
baseline, the preferred model, and the IDs from alert systems. Alerts
starting with MB and KB come from the MOA and KMTNet surveys. Events with
$^\dagger$ are present in the benchmark sample from Table~1.}  }
\end{landscape}
\newpage

\begin{figure}[h]
\centerline{
\includegraphics[width=6.4cm]{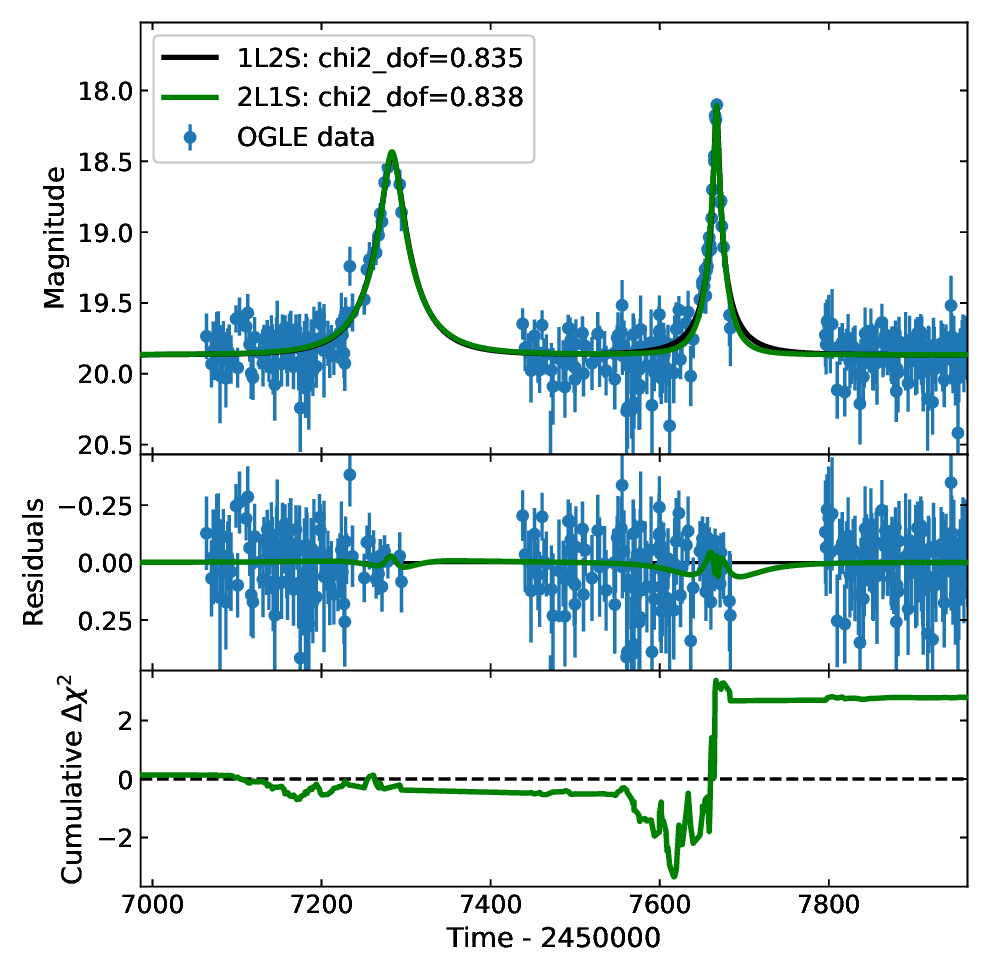}\hfill
\includegraphics[width=6.4cm]{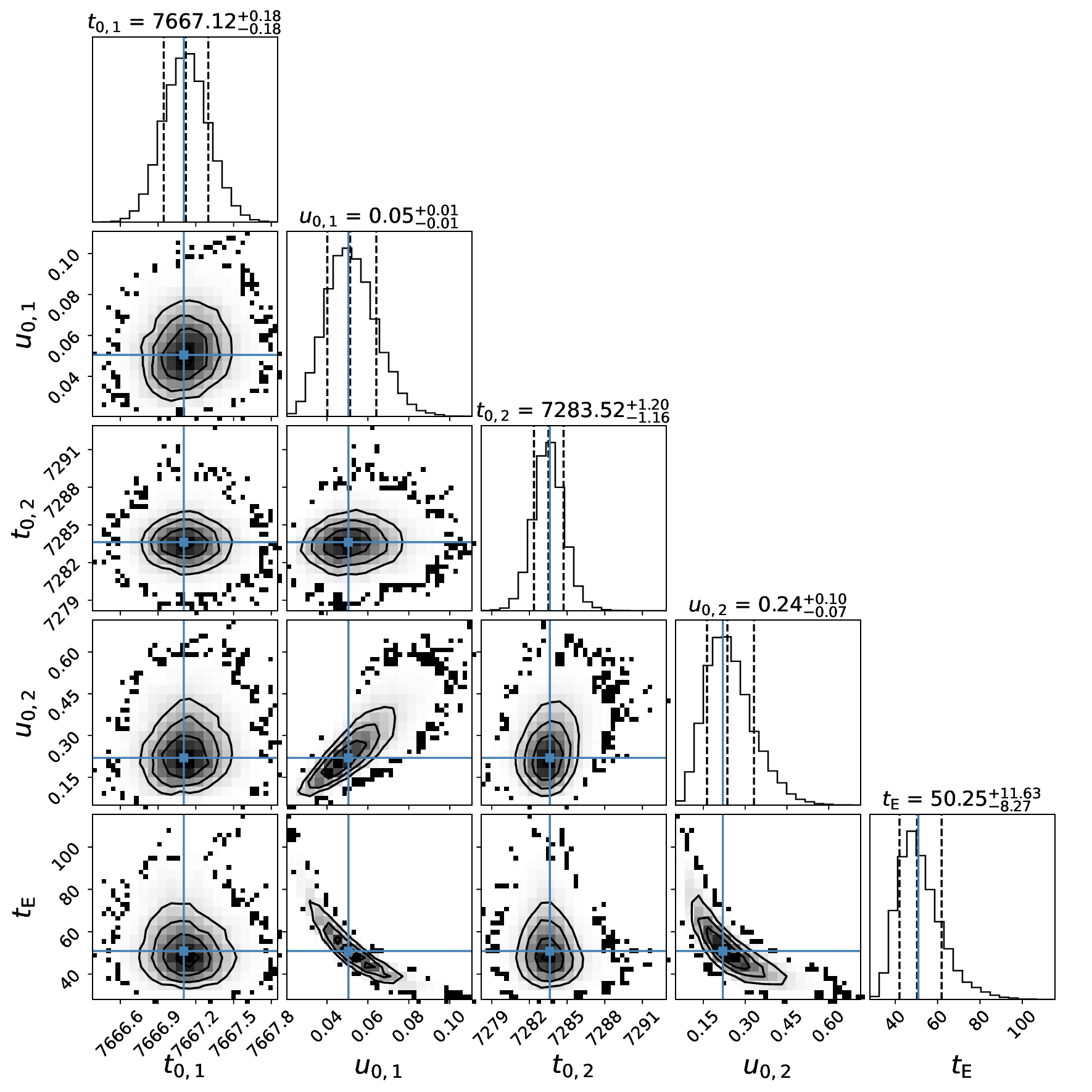}
}
\FigCap{Result for the event BLG615.27.39817 (alerted as OB151923 and
  KB160786), detected in a low-cadence field. {\it Left:} Light curve with
  the 1L2S and 2L1S models fitted, along with the residuals and cumulative
  $\Delta \chi^2$ distribution revealing the preference for the 1L2S
  model. {\it Right:} Corner plot with the posterior distribution and
  correlations for the five fitted parameters.}

\centerline{
\includegraphics[width=6.4cm]{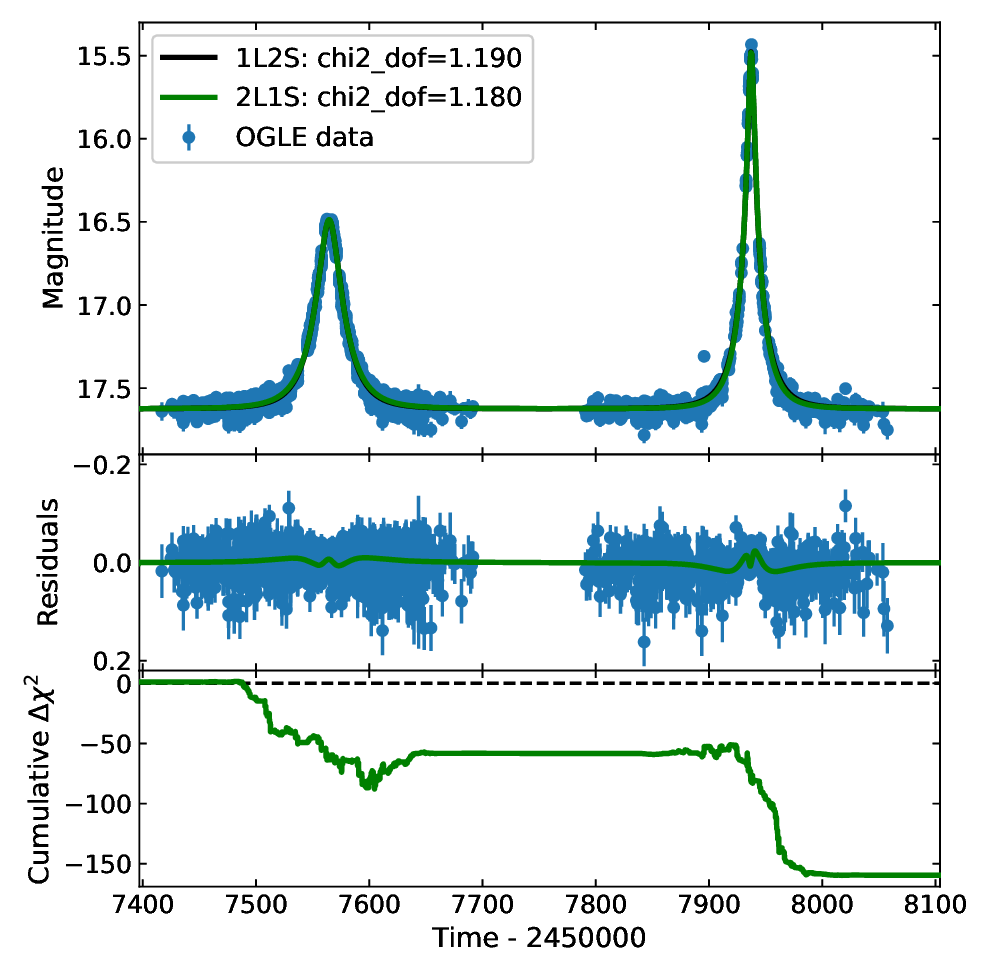}\hfill
\includegraphics[width=6.4cm]{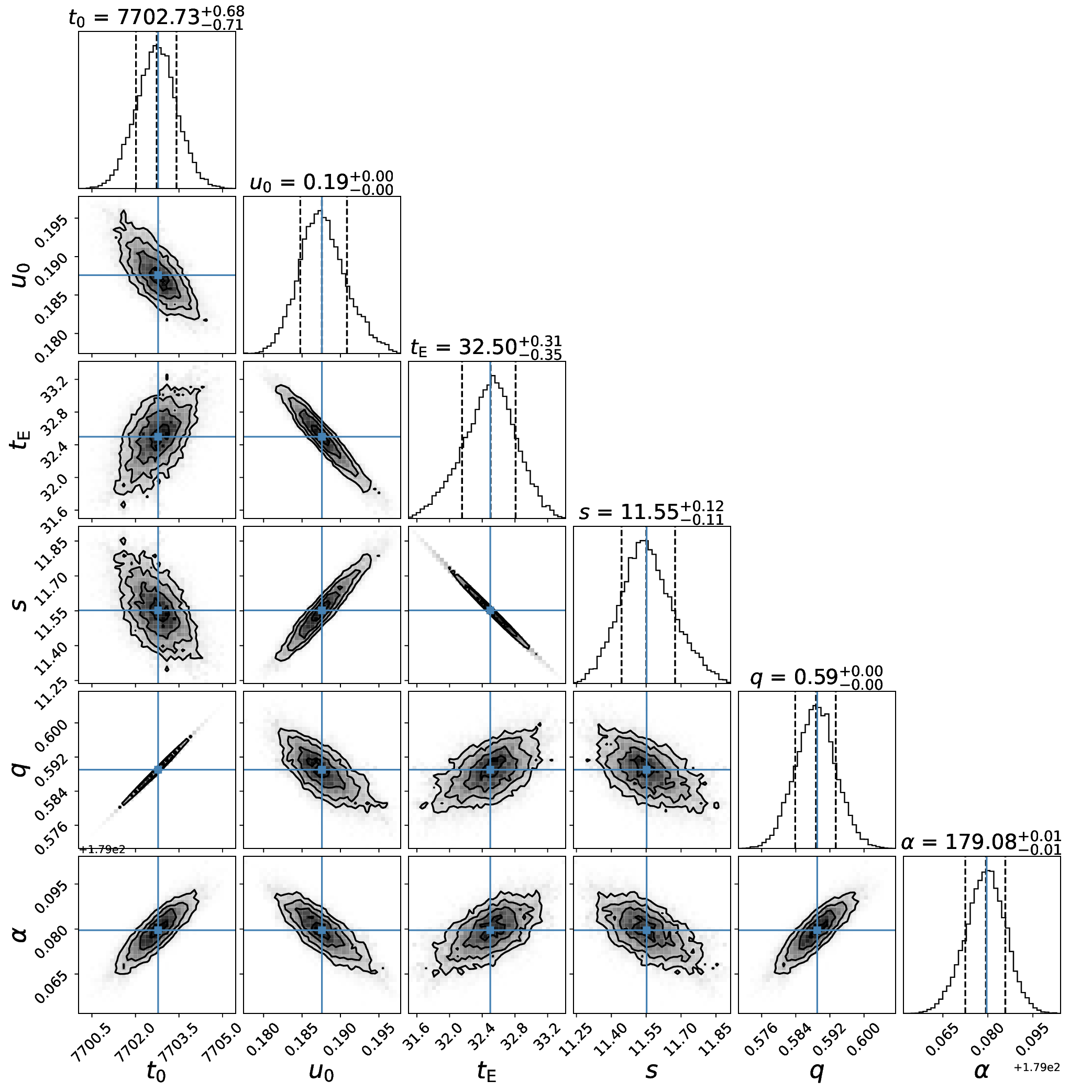}
}
\FigCap{Result for the benchmark event BLG505.03.213693 (alerted as OB161013, MB16309, and KB171620), with the 2L1S model
  more favored. The corner plots contain the posterior distribution of the
  six fitted parameters. Details as in Fig.~5.}
\end{figure}

Table~7 contains the results of the 1L2S fitting for all the detected events,
with the median and $1\sigma$ confidence interval values, blending parameters
for both sources ($f_{\rm s, 1}$, $f_{\rm s, 2}$), and $\chi^2_{\rm
fit}/\rm{dof}$. The $\chi^2_{\rm fit}/\rm{dof}$ value corresponds to the
model parameters with minimum $\chi^2$, which may differ from the given
median values if the posterior distributions are not symmetric. Table~8
gives the fitted 2L1S parameters, blending parameter $f_{\rm s}$, and
$\chi^2_{\rm fit}/\rm{dof}$ for all the events. The events BLG500.09.56443,
BLG500.31.22754, and BLG501.20.76926 are examples of derived $\chi^2_{\rm
fit}/\rm{dof}$ values for 1L2S and 2L1S equal to the third decimal place
and required further analysis of the evidence, blending flux, and $t_{\rm E}$
values.

\newpage
\begin{landscape}
\renewcommand{\arraystretch}{1.3}
  \MakeTableee{lccccccccc}{19cm}{Results of the binary source (1L2S) fitting
    for all the 107 detected events}
{\hline
\noalign{\vskip3pt}
Event      & $t_{0,1}$ & $u_{0,1}$ & $t_{0,2}$ & $u_{0,2}$ &
$t_{\rm E, 1L2S}$ & $f_{\rm s,1}$ & $f_{\rm s,2}$ &
$\chi_{\rm fit}^2/{\rm dof}$ & Pref.\\
& [HJD] & & [HJD] & & [d] \\
\noalign{\vskip3pt}
\hline
\noalign{\vskip3pt}
BLG500.04.66459 & $2458210.65^{+0.08}_{-0.08}$ & $0.026^{+0.011}_{-0.008}$ & $2458184.70^{+0.99}_{-1.03}$ & $0.168^{+0.114}_{-0.071}$ & $44.55^{+18.79}_{-12.55}$ & $0.09^{+0.04}_{-0.03}$ & $0.15^{+0.11}_{-0.06}$ & 1.065 &  \\ [1.5pt]
BLG500.09.56443 & $2456133.80^{+0.02}_{-0.02}$ & $0.079^{+0.028}_{-0.022}$ & $2456159.09^{+0.04}_{-0.05}$ & $0.117^{+0.084}_{-0.062}$ & $3.47^{+1.07}_{-0.71}$ & $0.21^{+0.07}_{-0.06}$ & $0.09^{+0.05}_{-0.03}$ & 0.918 & \checkmark \\ [1.5pt]
BLG500.22.46297 & $2456411.08^{+0.01}_{-0.01}$ & $0.064^{+0.001}_{-0.001}$ & $2457589.17^{+0.31}_{-0.29}$ & $0.099^{+0.006}_{-0.006}$ & $34.75^{+0.21}_{-0.20}$ & $0.734^{+0.005}_{-0.005}$ & $0.023^{+0.001}_{-0.001}$ & 0.672 &  \\ [1.5pt]
BLG500.27.59448 & $2455392.03^{+0.01}_{-0.01}$ & $0.011^{+0.003}_{-0.003}$ & $2455646.60^{+0.13}_{-0.11}$ & $0.019^{+0.016}_{-0.013}$ & $29.07^{+7.30}_{-5.94}$ & $0.10^{+0.03}_{-0.02}$ & $0.10^{+0.03}_{-0.02}$ & 0.913 & \checkmark \\ [1.5pt]
BLG500.31.22754 & $2458575.80^{+1.79}_{-1.89}$ & $1.145^{+0.093}_{-0.095}$ & $2456846.63^{+2.47}_{-2.41}$ & $1.317^{+0.115}_{-0.117}$ & $59.33^{+3.61}_{-3.09}$ & $0.70^{+0.17}_{-0.18}$ & $0.55^{+0.14}_{-0.15}$ & 0.994 &  \\ [1.5pt]
BLG501.08.186097 & $2456496.05^{+0.27}_{-0.27}$ & $0.876^{+0.153}_{-0.119}$ & $2457135.72^{+0.37}_{-0.36}$ & $0.898^{+0.156}_{-0.121}$ & $47.17^{+4.63}_{-4.70}$ & $0.24^{+0.08}_{-0.06}$ & $0.24^{+0.08}_{-0.07}$ & 0.664 &  \\ [1.5pt]
BLG501.13.138484 & $2458189.84^{+0.28}_{-0.25}$ & $0.124^{+0.029}_{-0.026}$ & $2455685.57^{+0.07}_{-0.07}$ & $0.132^{+0.029}_{-0.026}$ & $21.50^{+3.98}_{-3.04}$ & $0.15^{+0.04}_{-0.03}$ & $0.11^{+0.03}_{-0.02}$ & 0.969 & \checkmark \\ [1.5pt]
BLG501.13.151818 & $2456903.67^{+0.09}_{-0.09}$ & $0.086^{+0.004}_{-0.004}$ & $2457620.61^{+1.35}_{-1.30}$ & $0.308^{+0.025}_{-0.022}$ & $121.48^{+4.15}_{-3.96}$ & $0.113^{+0.005}_{-0.005}$ & $0.10^{+0.01}_{-0.01}$ & 1.062 &  \\ [1.5pt]
BLG501.14.93030 & $2455786.06^{+0.01}_{-0.01}$ & $0.002^{+0.002}_{-0.001}$ & $2457187.71^{+0.09}_{-0.09}$ & $0.169^{+0.032}_{-0.028}$ & $14.29^{+2.00}_{-1.65}$ & $0.14^{+0.02}_{-0.02}$ & $0.27^{+0.05}_{-0.04}$ & 0.937 & \checkmark \\ [1.5pt]
BLG501.20.76926 & $2458338.98^{+0.03}_{-0.03}$ & $0.537^{+0.020}_{-0.025}$ & $2458195.18^{+0.37}_{-0.37}$ & $0.806^{+0.064}_{-0.064}$ & $12.25^{+0.37}_{-0.28}$ & $0.81^{+0.06}_{-0.05}$ & $0.18^{+0.03}_{-0.02}$ & 0.726 &  \\ [1.5pt]
BLG501.22.125510 & $2458212.80^{+0.98}_{-0.97}$ & $0.285^{+0.112}_{-0.084}$ & $2458285.96^{+10.10}_{-12.64}$ & $1.113^{+0.312}_{-0.361}$ & $73.54^{+21.09}_{-14.21}$ & $0.52^{+0.84}_{-0.85}$ & $1.43^{+1.58}_{-2.04}$ & 0.883 & \checkmark \\ [1.5pt]
BLG501.30.90272 & $2456469.09^{+0.04}_{-0.09}$ & $0.060^{+0.061}_{-0.037}$ & $2456457.29^{+0.18}_{-0.18}$ & $0.611^{+0.266}_{-0.207}$ & $4.63^{+1.28}_{-0.92}$ & $0.11^{+0.08}_{-0.06}$ & $0.46^{+0.25}_{-0.25}$ & 1.089 & \checkmark \\ [1.5pt]
BLG501.31.52436 & $2455827.27^{+0.34}_{-0.32}$ & $0.686^{+0.028}_{-0.030}$ & $2456462.16^{+0.69}_{-0.67}$ & $1.400^{+0.052}_{-0.055}$ & $50.48^{+1.35}_{-1.21}$ & $0.60^{+0.06}_{-0.06}$ & $0.50^{+0.05}_{-0.05}$ & 0.724 & \checkmark \\ [1.5pt]
BLG502.22.138016 & $2457314.66^{+0.14}_{-0.15}$ & $0.084^{+0.016}_{-0.014}$ & $2457198.88^{+0.31}_{-0.29}$ & $0.109^{+0.021}_{-0.018}$ & $41.82^{+4.98}_{-4.29}$ & $0.77^{+0.12}_{-0.13}$ & $0.17^{+0.05}_{-0.04}$ & 0.993 &  \\ [1.5pt]
BLG503.25.47961 & $2456746.27^{+0.26}_{-0.26}$ & $0.453^{+0.078}_{-0.086}$ & $2456580.01^{+3.20}_{-2.47}$ & $0.828^{+0.140}_{-0.158}$ & $17.50^{+2.61}_{-1.67}$ & $0.58^{+0.18}_{-0.16}$ & $0.53^{+0.17}_{-0.15}$ & 0.913 &  \\ [1.5pt]
\noalign{\vskip3pt}
\hline
\noalign{\vskip3pt}
\multicolumn{10}{p{18.8cm}}{Table contains 15 first objects, all data
  available in the {\it Acta Astronomica Archive:} {\it
https://acta.astrouw.edu.pl/acta/2025/oli\_75\_1/table7.mrt}.\newline The parameters are
  given as the median and $1\sigma$ confidence interval values derived from
  the MCMC chains. The blending parameters $f_{\rm s, 1}$ and $f_{\rm s,
    2}$ are derived as $f_{\rm s, x}=F_{\rm S, x}/(F_{\rm S, 1}+F_{\rm S,
    2}+F_{\rm bl})$. The $\chi^2_{\rm fit}/\rm{dof}$ value corresponds to
  the model with lowest $\chi^2$, which is usually different from the given
  median values.  The last column shows if the binary source model is
  preferred over the binary lens.}}
\end{landscape}
\newpage
\begin{landscape}
\renewcommand{\arraystretch}{1.4}
  \MakeTableee{lccccccccc}{12.5cm}{Results of the binary lens (2L1S) fitting
    for all the 107 detected events}
{\hline
\noalign{\vskip3pt}
Event & $t_0$ & $u_0$ & $t_{\rm E,2L1S}$ & $s$ & $q$ & $\alpha$ & $f_{\rm s}$ & $\chi^2_{\rm fit}/{\rm dof}$ &Pref.\\
& [HJD] & & [d] & & & [\arcd] \\
\noalign{\vskip3pt}
\hline
\noalign{\vskip3pt}
BLG500.04.66459 & $2458189.82^{+1.25}_{-1.65}$ & $1.204^{+0.023}_{-0.033}$ & $24.70^{+1.33}_{-1.34}$ & $1.96^{+0.02}_{-0.01}$ & $0.013^{+0.002}_{-0.001}$ & $125.64^{+2.16}_{-1.53}$ & $3.21^{+0.39}_{-0.36}$ & 0.996 & \checkmark \\ [1.5pt]
BLG500.09.56443 & $2456142.65^{+1.57}_{-1.64}$ & $0.149^{+0.103}_{-0.040}$ & $2.97^{+0.52}_{-0.76}$ & $8.59^{+2.87}_{-1.23}$ & $0.54^{+0.16}_{-0.14}$ & $180.92^{+0.33}_{-0.30}$ & $0.34^{+0.19}_{-0.11}$ & 0.919$^{\dagger}$ &  \\ [1.5pt]
BLG500.22.46297 & $2456468.25^{+15.27}_{-9.95}$ & $0.048^{+0.012}_{-0.031}$ & $35.42^{+0.29}_{-0.20}$ & $37.21^{+0.97}_{-4.05}$ & $0.05^{+0.01}_{-0.01}$ & $179.47^{+0.40}_{-0.81}$ & $0.74^{+0.01}_{-0.01}$ & 0.660 & \checkmark \\ [1.5pt]
BLG500.27.59448 & $2455535.18^{+9.49}_{-7.79}$ & $0.043^{+0.045}_{-0.016}$ & $11.16^{+1.13}_{-2.15}$ & $22.86^{+5.46}_{-2.11}$ & $0.78^{+0.10}_{-0.11}$ & $0.37^{+0.09}_{-0.06}$ & $0.47^{+0.10}_{-0.09}$ & 0.917 &  \\ [1.5pt]
BLG500.31.22754 & $2457762.16^{+39.78}_{-31.34}$ & $1.113^{+0.078}_{-0.091}$ & $70.42^{+4.77}_{-3.76}$ & $24.59^{+1.39}_{-1.56}$ & $0.89^{+0.07}_{-0.08}$ & $359.58^{+0.14}_{-0.11}$ & $1.15^{+0.26}_{-0.26}$ & 0.994$^{\dagger}$ & \checkmark \\ [1.5pt]
BLG501.08.186097 & $2456812.31^{+1.51}_{-2.69}$ & $0.007^{+0.011}_{-0.005}$ & $62.49^{+6.11}_{-11.27}$ & $10.42^{+2.28}_{-0.92}$ & $0.98^{+0.01}_{-0.02}$ & $172.45^{+0.39}_{-0.83}$ & $0.29^{+0.19}_{-0.09}$ & 0.662 & \checkmark \\ [1.5pt]
BLG501.13.138484 & $2457069.49^{+29.63}_{-31.45}$ & $0.343^{+0.046}_{-0.032}$ & $11.72^{+0.63}_{-0.87}$ & $213.63^{+17.16}_{-10.92}$ & $0.81^{+0.04}_{-0.04}$ & $359.982^{+0.004}_{-0.004}$ & $0.73^{+0.08}_{-0.11}$ & 0.973 &  \\ [1.5pt]
BLG501.13.151818 & $2457296.68^{+16.24}_{-12.47}$ & $0.128^{+0.016}_{-0.011}$ & $168.25^{+1.19}_{-1.05}$ & $4.49^{+0.03}_{-0.03}$ & $0.82^{+0.06}_{-0.07}$ & $4.44^{+0.20}_{-0.16}$ & $0.123^{+0.004}_{-0.003}$ & 1.001 & \checkmark \\ [1.5pt]
BLG501.14.93030 & $2456973.18^{+14.09}_{-14.08}$ & $0.333^{+0.019}_{-0.021}$ & $7.67^{+0.13}_{-0.16}$ & $182.78^{+3.90}_{-3.11}$ & $0.18^{+0.01}_{-0.01}$ & $0.13^{+0.01}_{-0.01}$ & $0.86^{+0.05}_{-0.05}$ & 0.942 &  \\ [1.5pt]
BLG501.20.76926 & $2458296.08^{+2.78}_{-3.02}$ & $0.604^{+0.036}_{-0.037}$ & $13.94^{+0.56}_{-0.49}$ & $10.42^{+0.38}_{-0.40}$ & $0.42^{+0.04}_{-0.04}$ & $357.73^{+0.24}_{-0.27}$ & $0.90^{+0.08}_{-0.08}$ & 0.726$^{\dagger}$ & \checkmark \\ [1.5pt]
BLG501.22.125510 & $2458240.67^{+2.43}_{-2.35}$ & $0.603^{+0.089}_{-0.083}$ & $84.15^{+19.19}_{-14.16}$ & $1.43^{+0.12}_{-0.09}$ & $0.39^{+0.12}_{-0.10}$ & $189.95^{+3.40}_{-3.55}$ & $1.47^{+0.67}_{-0.91}$ & 0.883$^{\dagger}$ &  \\ [1.5pt]
BLG501.30.90272 & $2456459.99^{+0.62}_{-0.66}$ & $0.468^{+0.125}_{-0.085}$ & $4.92^{+0.52}_{-0.55}$ & $2.72^{+0.28}_{-0.21}$ & $0.27^{+0.09}_{-0.08}$ & $170.40^{+1.46}_{-1.97}$ & $0.51^{+0.10}_{-0.10}$ & 1.088$^{\dagger}$ &  \\ [1.5pt]
BLG501.31.52436 & $2456134.09^{+5.77}_{-5.92}$ & $0.899^{+0.059}_{-0.074}$ & $61.09^{+3.70}_{-2.45}$ & $10.49^{+0.43}_{-0.58}$ & $0.94^{+0.03}_{-0.03}$ & $183.29^{+0.12}_{-0.12}$ & $0.90^{+0.14}_{-0.13}$ & 0.724$^{\dagger}$ &  \\ [1.5pt]
BLG502.22.138016 & $2457290.65^{+2.67}_{-2.43}$ & $0.066^{+0.012}_{-0.008}$ & $42.19^{+1.42}_{-1.72}$ & $3.05^{+0.11}_{-0.08}$ & $0.26^{+0.03}_{-0.04}$ & $357.26^{+0.47}_{-0.32}$ & $0.73^{+0.04}_{-0.06}$ & 0.984 & \checkmark \\ [1.5pt]
BLG503.25.47961 & $2456646.86^{+10.78}_{-12.52}$ & $0.694^{+0.104}_{-0.107}$ & $22.50^{+3.41}_{-2.64}$ & $7.39^{+1.07}_{-1.10}$ & $0.64^{+0.22}_{-0.23}$ & $176.27^{+0.96}_{-1.52}$ & $1.00^{+0.24}_{-0.21}$ & 0.912 & \checkmark \\ [1.5pt]
\noalign{\vskip3pt}
\hline
\multicolumn{10}{p{18.8cm}}{Table contains 15 first objects, all data
  available in the {\it Acta Astronomica Archive:} {\it
    https://acta.astrouw.edu.pl/acta/2025/oli\_75\_1/table8.mrt}.\newline The
  $\chi^2$ values marked with $^{\dagger}$ are less than 0.001 different
  than the one derived in the 1L2S fit. These events required further
  analysis of the evidence, blending flux and $t_{\rm E}$ values, in order
  to determine which model is more representative.}  }
\end{landscape}

Out of 107 detected events, 59 resulted in more favored 2L1S models
whereas 48 resulted in 1L2S models. The close numbers demonstrate how
similar these models can be in the configuration of well-separated bumps
analyzed here. The evidence values derived using {\sf UltraNest} follow the
$\chi^2$ values in most cases, with very few cases having smaller evidence
and larger $\chi^2$ or vice-versa.

Analyzing the 1L2S and 2L1S separately, it is interesting to compare the
distribution of the parameters with physical relevance, such as the Einstein
timescale, projected separation, and the mass ratio (or, similarly, source
flux ratio). Fig.~7 presents histograms with the mentioned parameters,
with the results from the preferred model for all binary events. The $t_{\rm
E}$ histograms closely follow the adopted prior from Mróz \etal (2020),
with peaks around 35--40~d. The distribution for the separation $s_{\rm
1L2S}$ has a higher median value and larger upper uncertainties than for
2L1S. The distributions for source flux ratio and mass ratio do not present
a peak and are virtually flat between 0 and 1.

The largest binary lens separation in literature of $s\approx17$ was reported
by Poleski \etal (2014). This separation was between two stars of which one
hosts a planet. The literature studies of larger number of events by Skowron
\etal (2009) and Jaroszyński \etal (2010) found events with largest
separations of $s=6.57$ and 4.08, respectively. In the present work, the
largest separation values were derived for the binary lens event
BLG500.22.46297 ($s\approx35$) and the binary source events BLG501.13.138484
and BLG501.14.93030 ($s_{\rm 1L2S}\approx100$).

Almost none of the detected events have previous work in the literature, but
some have data in other observational projects. For example,
BLG504.06.110803, BLG505.30.204598, and BLG523.23.114690 have data from K2
Campaign 9 of the Kepler mission (K2C9, Henderson \etal 2016).
Similarly, four events have Spitzer data (\eg Yee \etal 2015, Calchi
Novati \etal 2015): BLG504.06.184302, BLG512.14.160752, BLG534.18.5145, and
BLG614.31.14827.

\begin{figure}[htb]
\centerline{\includegraphics[width=12.5cm]{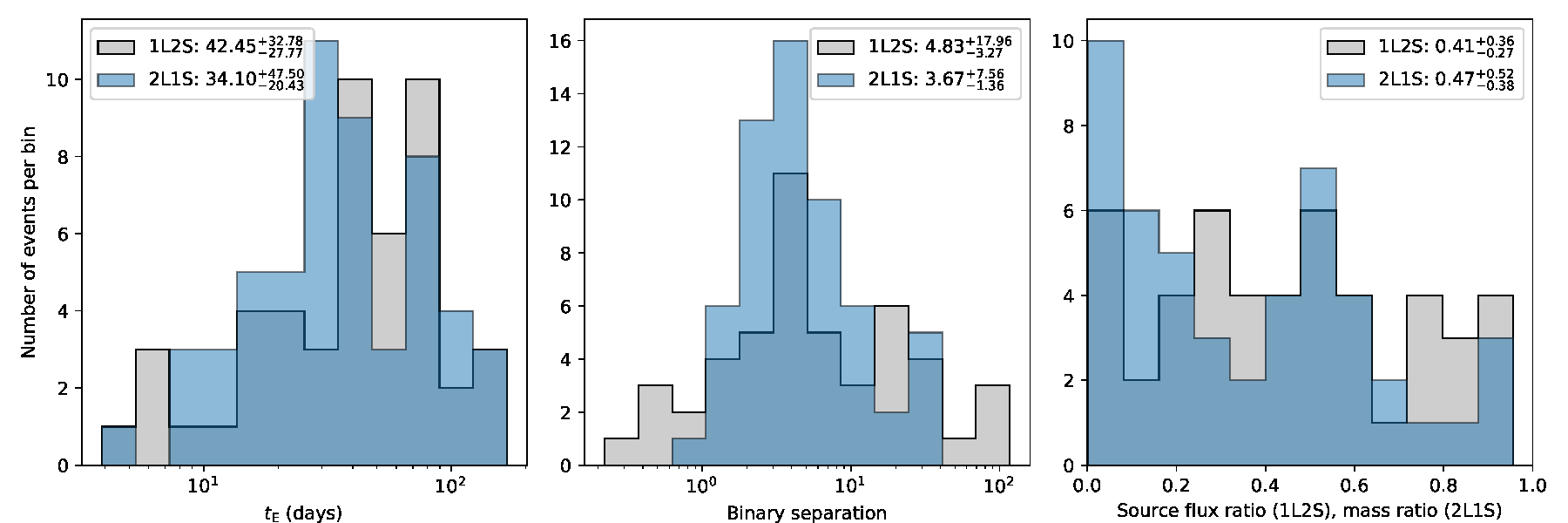}}

\FigCap{Comparison of the physically relevant parameters derived for binary
  source and binary lens models, namely $t_{\rm E}$, separation, and source
  flux ratio (1L2S) or mass ratio (2L1S). The median values and $1\sigma$
  confidence intervals are given in the labels. The binary sources are
  probably composed of two giant stars or giant + main sequence star, for
  which the relation $ \propto M^{3.5}$ does not hold.}
\end{figure}

\Section{Summary}
Several studies with large samples of microlensing events do not include
binary events and apply a correction to account for them. The proper
assumption of these events and the derivation of binary fraction and mass
ratio distributions allow to derive the bulge IMF from the observed
luminosity function (Calamida \etal 2015), among other possibilities.

In this work, we adapted an algorithm that searches for microlensing events
in the OGLE-IV database (Mróz \etal 2017, 2019), in order to detect binary
events with two well-separated bumps. The model fitting was carried out with
{\sf MulensModel} (Poleski and Yee 2019), in order to identify if the event
is better represented by a binary source or binary lens model. A series of
criteria were applied which resulted in a sample of 107 binary events, with
very good fitted models.

Most of the detected events were alerted by the OGLE, MOA, and/or KMTNet
systems, which enhances the robustness of the results. A few dozen events
had very similar $\chi^2_{\rm fit}/\rm{dof}$ values between the binary source
and binary lens models, requiring a more careful analysis of the Bayesian
evidence, the amount of negative blending flux, and the derived $t_{\rm E}$
compared to the adopted prior from Mróz \etal (2020).

A visualization of all results together showed that the $t_{\rm E}$
distribution peaked around 35--40~d for both binary source and binary lens
events, in accordance with the prior results from previous PSPL
analysis. The derived separations between the sources and the lenses peak
around 4.5 and 3.7, respectively, with a broader distribution for binary
sources.  The mass ratio and source flux ratio follow a virtually flat
distribution between 0 and 1.

The derived mass ratio for the binary lens models is of fundamental
importance for the future study of the IMF. For the binary source models,
the source flux ratio and the evolutionary stage of the source star will be
assumed to obtain information about the mass.  Additionally, corrections for
detection efficiency will be applied to account for events that were skipped
in the new set of thresholds and other changes in the method.

The tools developed in this work for both event detection and dealing with
the degeneracy between binary lens or binary source models may be valuable
for future analysis of data from the Roman Space Telescope (Penny \etal
2019). The binarity statistics and IMF derived in the following works can
also be used to measure the mass and distance of lenses.

\Acknow{This research was funded in part by the National Science Center,
  Poland, grant Sonata Bis 2021/42/E/ST9/00038 to Radoslaw Poleski. For the
  purpose of Open Access, the author has applied a CC-BY-4.0 public
  copyright license to any Author Accepted Manuscript (AAM) version arising
  from this submission.}

\end{document}